\documentclass[nofootinbib,aps,10pt,twocolumn]{revtex4-1}
\usepackage{amstext,amssymb}
\usepackage{amsmath}
\usepackage{graphicx}
\usepackage[hyperfootnotes=true]{hyperref}
\usepackage{color}
\usepackage{comment}
\usepackage{float}
\usepackage{array,multirow}
\usepackage[most]{tcolorbox}
\usepackage{slashed} 
\usepackage{multirow}
\usepackage[T1]{fontenc}
\usepackage[format=hang,font={normalsize,sf},labelfont=bf,justification=raggedright]{caption}
\usepackage{cancel}

\usepackage{slashed}

\def \epsilon {\varepsilon} 

\def \vec#1{{\boldsymbol{#1}}}

\begin{document}
\title{\bf Numerically analyzing self-interacting dark matter}
\author{Utkarsh Patel$^a$}
\email{utkarshp@iitbhilai.ac.in}
\author{Sudhanwa Patra$^a$}
\email{sudhanwa@iitbhilai.ac.in}
\affiliation{$^a$Department of Physics, Indian Institute of Technology Bhilai, Raipur-492015, India\\}

\begin{abstract}

We consider the scenario of self-interacting dark matter(SIDM) with a light mediator in a model independent way, which can alleviate two long standing issues of the small scale cosmology namely cusp vs. core and too-big-to-fail. A Yukawa potential is chosen to achieve mediator exchange between DM particles as part of their self-interactions. The dynamics of self-interacting transfer cross-section is studied for a range of mediator mass($m_Z'$). Also, a relationship is established between the cross-section and DM particles' relative velocity, which ensures the solution to DM crisis at small scales. Our obtained numerical results are efficient compared to the earlier works in the context that lesser number of $\ell$ modes have been used by us to achieve the same level of accuracy in the cross-section calculations. For a better understanding of the SIDM parameter space, we perform an analytical analysis on the dependence of transfer cross-section over the other important SIDM parameters using a Hulth\'{e}n potential which is similar in it's behaviour to Yukawa potential. A detailed evolution of particle dynamics using the Boltzmann equation and the effect of sommerfeld enhancement on such calculations has also been studied here. We also provide a minimal anomaly-free leptophilic extension of standard model, that can incorporate SIDM and its mediator candidate in the framework.
\end{abstract}

\maketitle
\section{Introduction}
Dark Matter\cite{Zwicky:1933gu,Rubin:1980zd} today remains as one of the many unsolved mysteries in the field of physics. There have been many theories or models put forward by great minds to explain its existence and effects on our understanding of the universe but the results of no such single theory have been experimentally verified with a 5$\sigma$ confidence level. The current best theory available in the field of cosmology is the $\lambda$-CDM model of the universe\cite{Bahcall:1999xn}. Here, CDM is the Cold Dark Matter which assumes Dark Matter particles to have non-relativistic velocities with no self-interactions\cite{Peebles:1982ff}. Simulated results using $\lambda$-CDM matches well with the observational data of the structures at the large scales of the universe\cite{Springel:2006vs}. However, it is still far from clear that CDM paradigm can also successfully account for small scale structures. CCDM( collisionless CDM) predicts cuspy profiles\cite{Navarro:1996gj} for galaxies which is in contrast with the observational data from dwarf galaxies showing DM distribution with cores\cite{Oh:2010ea}. Other issues like too big to fail\cite{Boylan-Kolchin:2011qkt,Boylan-Kolchin:2011lmk}, missing satellite problem\cite{Klypin:1999uc,Kauffmann:1993gv,Moore:1999nt}, diversity problem\cite{Oman:2015xda} have also put a question mark on the success of CCDM at small scales. Although a counter-argument in favor of CCDM, pointing to a lack of enough computational power to numerically simulate together the baryonic and non-baryonic matter of the galaxies and thus giving us a better picture of galactic structure, could be put forward\cite{Governato:2009bg,Navarro:1996bv,Oh:2015xoa,Oh:2010mc,Frusciante:2012jg,Kaplinghat:2019dhn}.
\par Instead of waiting for years for numerical simulations to be more computationally complex, researchers have put forward alternate theories to tackle the challenges faced by CCDM models. Early attempts of such type supposed the DM particles to be quasi-relativistic during kinetic decoupling from the thermal bath in the early universe, which simply means the particles were assumed to be warm instead of cold\cite{Colombi:1995ze,Bode:2000gq}. Another promising alternative, which is also the main topic of focus for this paper is the self-interacting dark matter(SIDM) paradigm\cite{Spergel:1999mh}. In this scenario, DM particles scatter elastically among each-other, which leads to radical deviation from the CDM predictions for the small scale structures. Also, the scattering rate for DM self interactions is proportional to the DM density, giving a similar structure formation at large scales with that of the CDM model.
\par Based on the simulation data, it is seen that the typical self interacting cross-sections needed to resolve core-cusp and other small scale issues is $\mathbf{\sigma\sim 10^{-24}{\text{\textbf{cm}}}^2\times (m_{\chi}/\text{\textbf{GeV}})}$\cite{Vogelsberger:2012ku,Rocha:2012jg,Zavala:2012us}. Here, $m_{\chi}$ is the DM mass. Also, observations from astrophysical data provide constraints on this cross-section. As an example, the findings from Bullet cluster observations provide a bound as $\sigma_T/m_{\rm DM} < 1.25$ $\hbox{cm}^2/\hbox{g}$ at 68$\%$ CL \cite{Clowe:2003tk,Markevitch:2003at,Randall:2008ppe} and some other cluster merger observations gives $\sigma_T/m_{\rm DM} < 0.47$ $\hbox{cm}^2/\hbox{g}$ at 95\% CL \cite{Harvey:2015hha}. Such large self-interacting cross-sections of DM naturally come out in scenarios where DM has a light mediator\cite{Spergel:1999mh,Feng:2009mn,Ackerman:2008kmp}. One remarkable feature of such SIDM models, if potential is carefully chosen, is that the interaction cross-section of DM particles can be made to depend on DM velocities. The motivation for this is straight-forward. Given a velocity dependence, cross-section is stronger at small-scales and is reduced at large scales(due to large velocities of DM), thus not only solving the small scale issues but also to remain consistent with CDM predictions at large scales\cite{vandenAarssen:2012vpm,Bringmann:2016din,Kaplinghat:2015aga,Tulin:2013teo}. The Yukawa potential can lead to rich dynamics and also provide for a velocity dependent cross-section. 
 
\par Several previous works have explored the light mediator scenario of SIDM within various BSM models\cite{Kouvaris:2014uoa,Bernal:2015ova,Kainulainen:2015sva,Hambye:2019tjt,Cirelli:2016rnw,Kahlhoefer:2017umn,Dutta:2021wbn}. The self interactions can be mediated either via scalar or vector mediators, but are more naturally explained through vector mediators, as they are easily realized within gauge extensions of the SM. As an example, in the recent work \cite{Borah:2021pet}, authors consider the SM extended with a $U(1)_D$ symmetry, consisting of a vector like fermion DM charged under the extended gauge group. Here, $Z'$ vector boson acts as the mediator particle for self-interactions of DM. Other similar works include \cite{KumarBarman:2018hla,Kamada:2018kmi,Balducci:2018ryj,Kamada:2018zxi,Lee:2020eap,Ho:2022erb,Heeck:2022znj}. Based on the similar line of work, we here firstly study the general model independent approach to perform calculations realted to SIDM methodology. We also show a way to enhance the numerical calculations performed in reference~\cite{Tulin:2013teo} so that the desired level of accuracy in results is achieved by summing over lesser $\ell$ modes. At the end, we present a minimal extension of SM to include a gauge group $U(1)_{\ell}$, where $\ell$ denotes lepton number. Here, a Dirac fermion charged under $U(1)_{\ell}$ acts as DM. The model has light mediator interactions in terms of $U(1)_{\ell}$ gauge boson $Z'$, required for SIDM phenomenology. Our focus in this work is to study the self-interacting DM cross-section and it's dependence on mediator mass, dependence of DM cross-section on DM velocity, relic abundance of DM and direct \& indirect detection bounds based on the coupling of the dark matter particle to the $Z'$. Future prospects of the work would include a detailed study on the $U(1)_{\ell}$ extension model, collider constraints and experimentally verifiable parameters of the model.

In this work, we focus on the interesting case of self-interacting dark matter in a light mediator scenario, the results of which have important implications on the structure formation and evolution. Firslty, we present the current status of research in the field including the established observational and theoretical results. In the next section, we focus on the general model-independent approach studying the basic methodology required in a SIDM framework. In the third section, a Hulth\'{e}n potential is chosen in place of the yukawa potential as it allows an analytical solution for s-wave mode in the partial wave analysis, thus enabling us to verify our numerical results. In the fourth section, the constraints coming from relic density observations on the parameter space for SIDM are studied including the sommerfeld effects. In the last section, we provide a simple extension of SM to include a leptophilic symmtery which can serve as a potential minimal framework to provide for DM and it's mediator candidates.

\section{General framework for SIDM}
We here consider a hidden sector with $Z'$ portal for studying self interactions of Dark Matter. This could be realised in a simple $U(1)$ extension of Standard Model via the kinetic mixing between the hidden and visible sectors with the following interaction term: 

\begin{eqnarray}
\mathcal{L}&=& \mathcal{L}_{SM}+ i\bar{\chi}\slashed{D}\chi - m_{\rm DM}\bar{\chi}\chi-\frac{1}{4} Z'^{\mu\nu} Z'_{\mu\nu} \nonumber \\
&&-\frac{\epsilon}{2} B^{\mu\nu} Z'_{\mu\nu} +\frac{1}{2}m_{Z'}^{2} Z'^{\mu} Z'_{\mu} + \ldots,
\label{eq:lag_km}
\end{eqnarray}

Here, $\chi$ is the dark matter candidate, $Z'$ is the corresponding mediator particle and $\epsilon$ is the strength of kinetic mixing parameter. The covariant derivative is given by $D_{\mu}=\partial_{\mu}+ig_{\chi}Z'_{\mu}$, where $g'{\chi}$ is the gauge coupling and $Z'$ is the corresponding gauge boson of $U(1)$ gauge group. The interaction resulting between dark matter particle and the mediator $(g_{\chi}\overline{\chi}\gamma^{\mu}\chi Z'_{\mu})$ plays the key role in defining the self interaction cross-sections of DM in the dark sector, which in turn seems to resolve the small-scale issues.\cite{Vogelsberger:2012ku,Rocha:2012jg,Zavala:2012us}. While on the other hand, at large scale, the kinetic mixing ensures the correct relic density of DM via $\sigma v(\chi\chi\rightarrow f\overline{f})$. 
\par The stability of the DM within this hidden sector is ensured by the suitable U(1) charges of the Dark Matter. The mass of the new gauge boson can arise through standard Higgs mechanism\cite{Englert:1964et,Higgs:1964pj} or via the St\"uckelberg mechanism\cite{Stueckelberg:1938hvi}. This topic is dealt with in details along with the introduction of a model framework, in the subsequent sections.

\begin{figure}[t!]
\begin{center}
 \includegraphics[trim={.08cm 0 0 0},clip,scale=0.95]{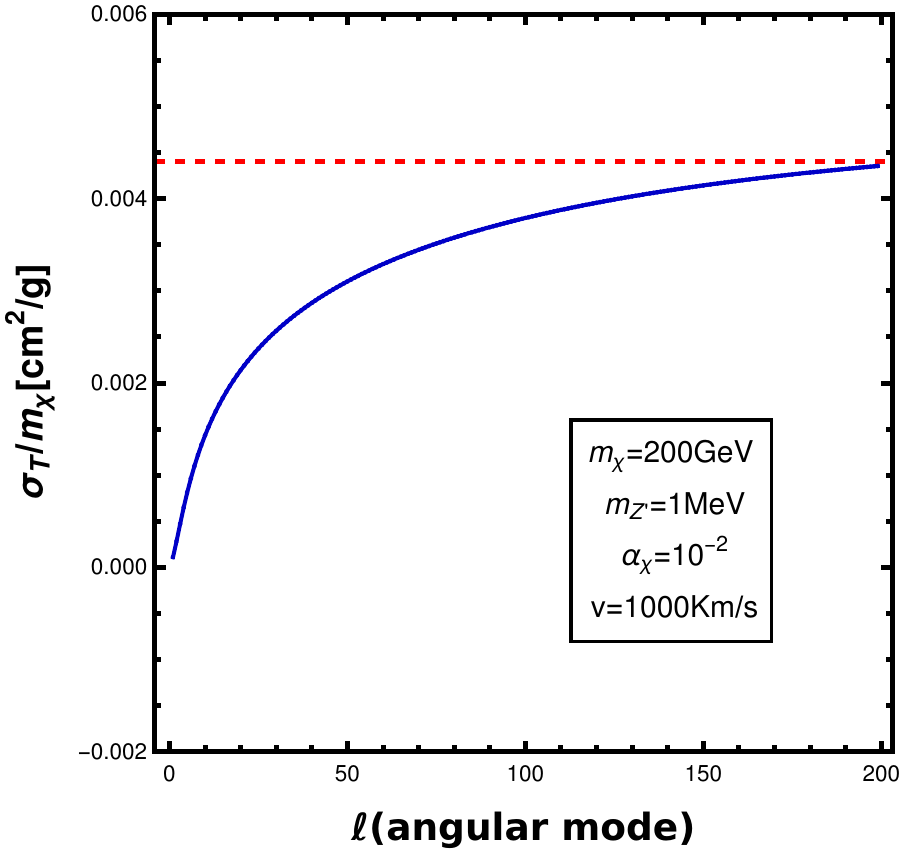}
\end{center}
\vspace{-0.25in}
\caption{Numerical calculation of $\sigma_T/m_{Z'}$(blue curve) depicting a saturation to classical analytical result(dotted red line) with increasing $\ell$ modes.}
\label{fig:lmode}
\end{figure}
\subsection{DM self interaction cross-section}
The required value of self interaction cross-section $0.10 \text{ cm}^2/\text{g} < \sigma_T/m_{\rm DM} < 10 \text{ cm}^2/\text{g}$ at dwarf scale, is around $13-15$ orders of magnitude larger than the typical WIMP scale cross-section\cite{Vogelsberger:2012ku,Rocha:2012jg,Peter:2012jh,Zavala:2012us}. It is already known that standard freeze-out annihilation or production channels fail to provide such large cross-section values. Thus, a new gauge portal interaction involving a light mediator could serve the purpose. The key parameters on which the cross-section value of SIDM can depend on, are:
$$ m_{\chi}, m_{Z'}, \alpha_{\chi}, v$$ Here, $\alpha_{\chi}= \frac{g_{\chi}^2}{4\pi}$ and $v$ is the DM particle velocity.
The details of this cross-section are governed by non-perturbative calculations. Also, we are required to have a interaction cross-section whose strength die out as we move towards the larger scales of the universe, because non self-interacting WIMP dark matter simulation results matches excellently with the observations at large scales\cite{Bahcall:1999xn,Springel:2006vs,Trujillo-Gomez:2010jbn}.  The simplest motivation for such a cross-section can be drawn from the nucleonic interactions, where the interaction cross-section is velocity dependent\cite{Chadwick:2011xwu}. In such scenarios, Sommerfeld effect plays a vital role with the idea that the wave-function of the DM gets modified due to multiple exchange of light mediators before the main interaction, leading to large enhancment of the scattering cross-section. This mediator exchange ,as motivated from nucleonic interactions, is realised by a Yukawa-type potential, given as:
\begin{equation}
 V(r)=\pm \frac{\alpha_{\chi}}{r}e^{-m_{Z'}r}
\end{equation}
Here, the potential is assumed to be spherically symmetric.

\begin{figure*}[t!]
 \includegraphics{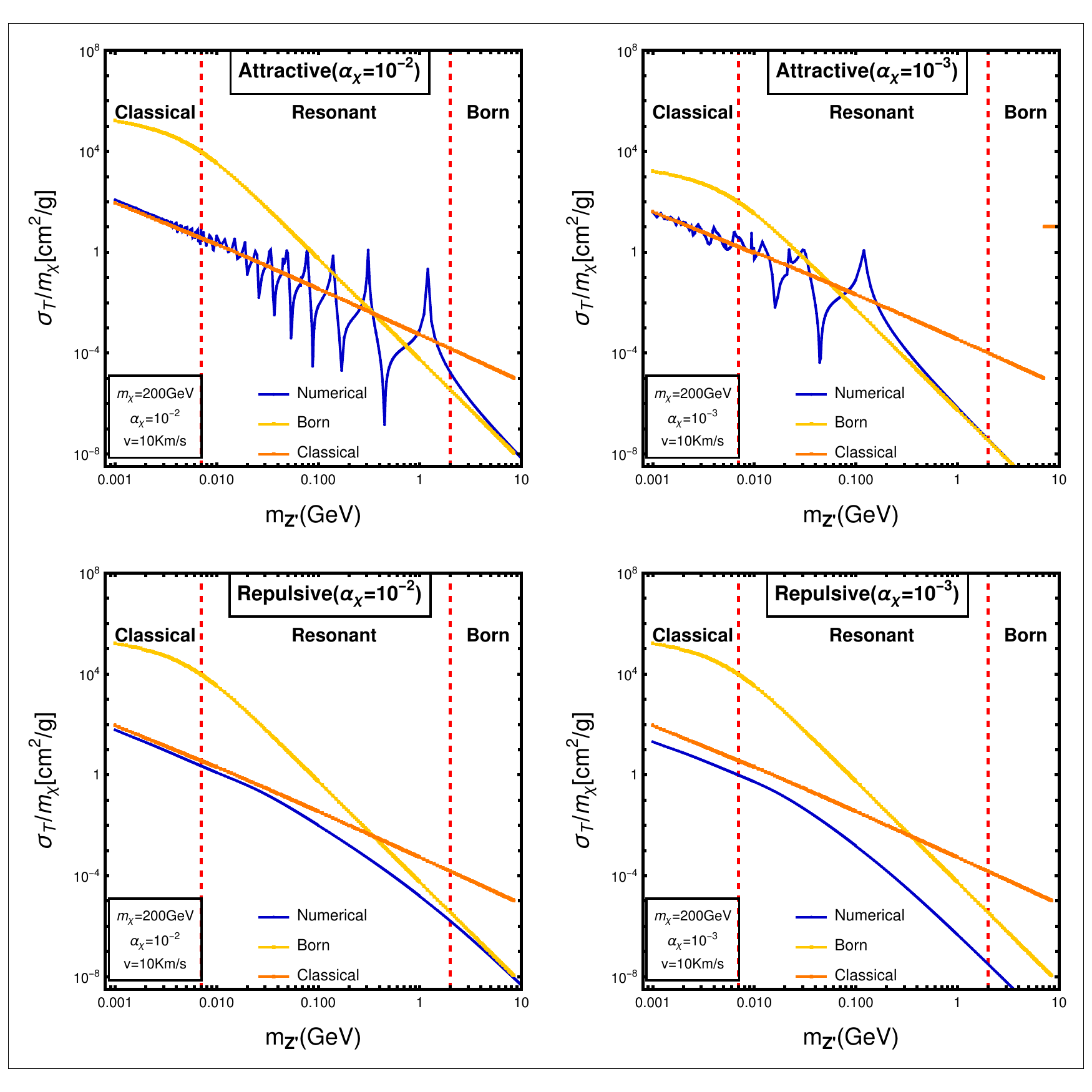}
\vspace{-0.15in}\caption{Figure shows the relation between transfer cross-section per DM mass and the DM mediator mass calculated using the numerical formula(Blue Plot). Also, the same is compared with the analytical results obtained using Born approximation(Yellow Plot) and the Classical non-perturbative formula(Orange Plot). The above two plots are obtained for attractive potential case with different coupling strengths. The other two plots are for a repulsive Yukawa-type potential. The formation of resonant peaks and valleys in the attractive case is clearly visible.}
\label{fig:sigmamphi}
\end{figure*}

\par In the simple 2-particle scenario, the scattering of DM particles and their wave-function behaviour can be studied by solving the Schrodinger equation in the presence of external potential. For the purpose of this work, Yukawa-type potential is chosen. The usual parameter to focus is the differential scattering cross-section $\frac{d\sigma}{d\Omega}$, however if the mediators are light($\lesssim1GeV$), then plasma literature dictates that the more relevant quantity is the transfer cross-section\cite{PhysRevA.60.2118}, which is the usual differential cross-section weighted by a $(1- \cos\theta)$ parameter. Thus, transfer cross-section is given as:
\begin{equation}
\boxed{
 \sigma_T=\int\frac{d\sigma}{d\Omega}(1- \cos\theta) d\Omega}
\end{equation}
Here, $\theta$ is the scattering angle.   

This transfer cross-section$(\sigma_T)$ when computed perturbatively(i.e. in the Born approximation) keeping only the dominant t-channel contributions in the interactions, is of the form\cite{Feng:2009hw}:
\begin{equation}
 \sigma_T= \frac{2\pi}{m_{Z'}^2}\beta^2\Bigg[\ln(1+R^2)-\frac{R^2}{1+R^2}\Bigg]
 \label{eq:eqn4}
\end{equation}
Here, $\beta=2\alpha_{\chi}m_{Z'}/(m_{\chi}v^2)$ and $R\equiv m_{\chi}v/m_{Z'}$. This above equation~\ref{eq:eqn4} works well only within the limit $\alpha_{\chi}m_{\chi}/m_{Z'}\ll1$\textbf{(Born Regime)}. 

\begin{figure*}[t!]
\begin{center}
\includegraphics[scale=0.4]{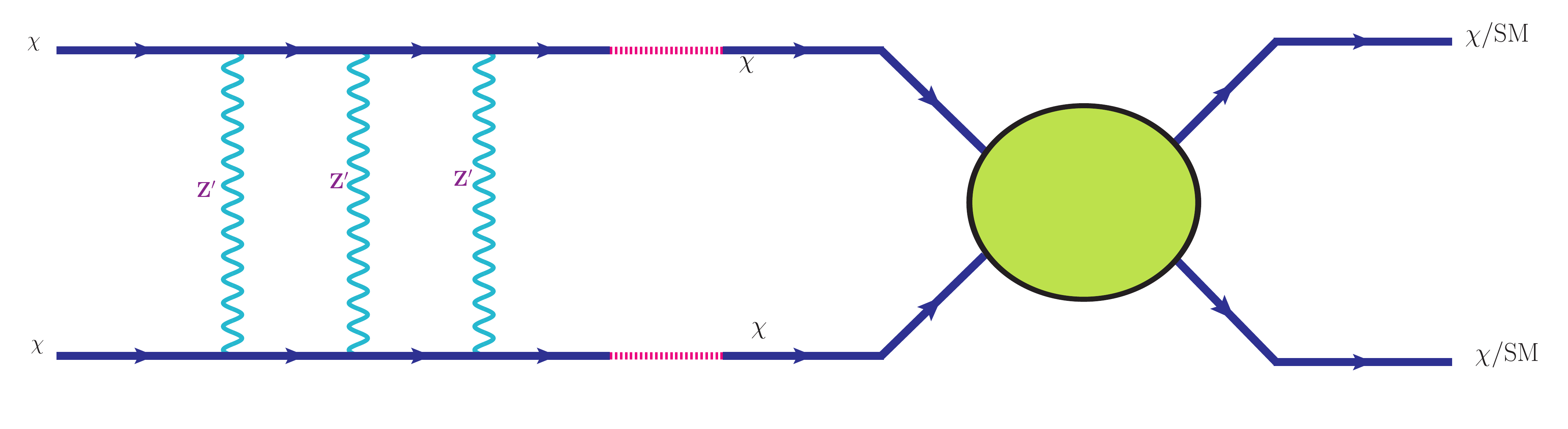}
\end{center}
\caption{Sommerfeld effect in SIDM($\chi$) via a light mediator.}
\label{fig:sommerfeld}
\end{figure*}

The parameter space where Born approximation loses validity is characterized by non-perturbative corrections. In here, a considerable amount of interactions occur outside the Debye sphere and thus they provide a substantial amount of contribution to the transfer cross-section. Within this regime, analytical formula has been derived in the literature only for the cases where quantum mechanical effects are sub-dominant~\cite{1308514}. This is marked by the condition, $m_{\chi}v/m_{Z'}\gg1$\textbf{(Classical Regime)}. For an attractive Yukawa-type potential, the expressions for $\sigma_T$ for various $\beta$ ranges as given in~\cite{1308514} are:

\begin{equation}
\sigma_T = 
\left\{\begin{array}{lc}
\frac{4 \pi}{m_\phi^2} \beta^2 \ln\left(1+\beta^{-1}\right) & \beta \lesssim 10^{-1} \\
\frac{8 \pi}{m_\phi^2} \beta^2 / \left(1+1.5 \beta^{1.65}\right) & \; 10^{-1} \lesssim \beta \lesssim 10^3 \\
\end{array} \right. \label{plasma} \, ,
\end{equation}

\begin{figure*}[t!]
\includegraphics[scale=0.9]{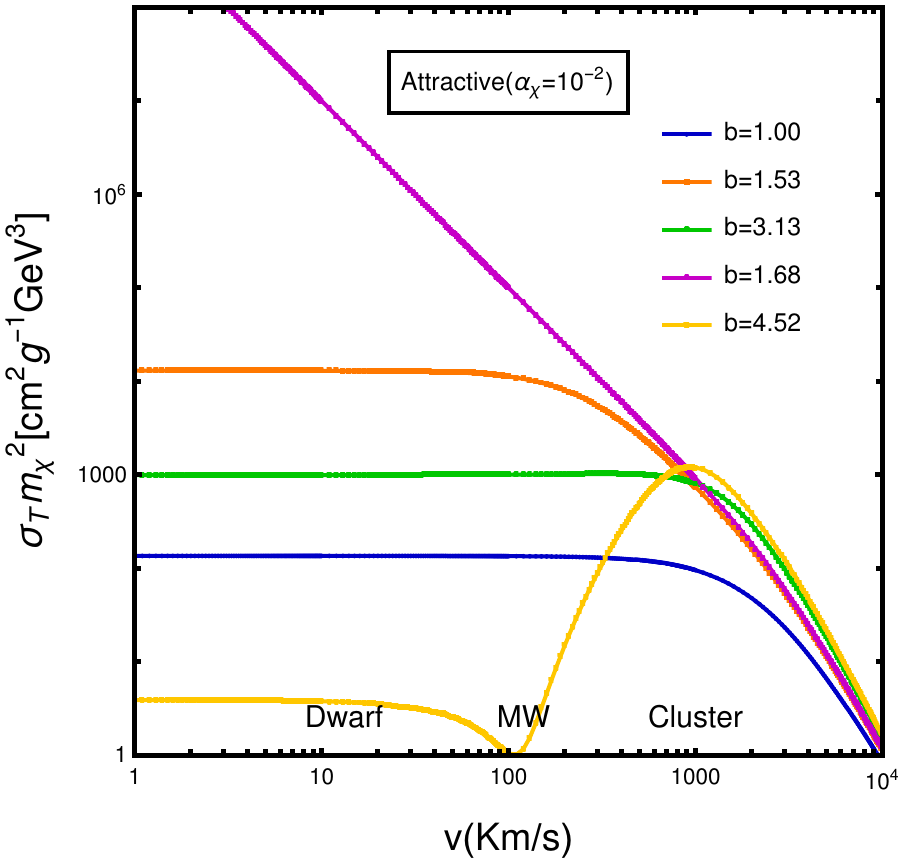}
\includegraphics[scale=0.9]{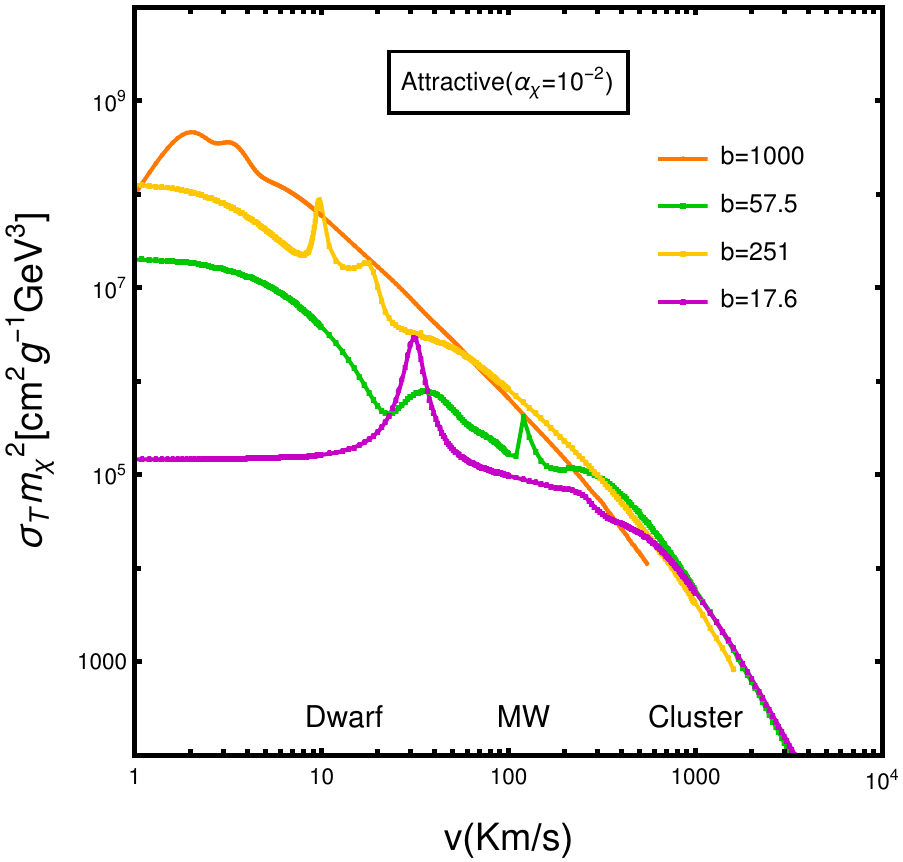}
\caption{Velocity dependence of $\sigma_T$ for different values of parameter "b" using the numerical solution.}
\label{fig:velnum}
\end{figure*}

Upon analyzing the expressions in \ref{eq:eqn4} and \ref{plasma}, it can be seen that a vast amount of parameter space remains unexplored. This region can be expressed as being  goverened by both \textbf{non-perturbative} and \textbf{quantum mechanical} effects. The authors in~\cite{Tulin:2013teo} have focused broadly on this region, which they call as \textbf{"resonant regime"} indicating the occurrence of quantum mechanical resonances in the value of $\sigma_T$ due to the Sommerfeld effect from the potential. There is no analytical expression available in the literature for $\sigma_T$ in this parameter space region. Thus, the only possibility is to numerically solve the Schrodinger equation to obtain the transfer cross-section.
\par From the theory of quantum scattering, the differential scattering cross-section is given as:
\begin{equation} \label{diffsigma}
\frac{d \sigma}{d\Omega} = \frac{1}{k^2} \Big| \sum_{\ell = 0}^\infty (2 \ell + 1) e^{i \delta_\ell} P_\ell(\cos\theta) \sin \delta_\ell \Big|^2 \, ,
\end{equation}

Here, $\delta_\ell$ is the phase shift for a partial wave $\ell$. The total scattering cross-section calculated from equation~\ref{diffsigma} is given by the sum over all angular momenta $\ell$ as:
\begin{eqnarray}
\sigma &=& \int d\Omega ~\frac{d\sigma}{d\Omega} \nonumber \\
&=& \frac{4\pi}{k^2}\sum_{\ell = 0}^{\infty} (2\ell + 1) \sin^2 \delta_{\ell}  \,.
\label{sigmaTsum}
\end{eqnarray}
Applying the same treatment for a weighted differential scattering cross-section i.e. 
\begin{eqnarray}
\sigma_{T} &\equiv& \int d\Omega (1-\cos\theta) \frac{d\sigma}{d\Omega} \nonumber \\
&=& \frac{4\pi}{k^2}\sum_{\ell=0}^{\infty} [  (2\ell +1)\sin^2\delta_\ell \nonumber \\
&~& -2(\ell+1) \sin \delta_\ell \sin\delta_{\ell+1} \cos(\delta_{\ell+1}-\delta_\ell)] \nonumber \\
&=& \frac{4\pi}{k^2}\sum_{\ell = 0}^{\infty} (\ell + 1) \sin^2 (\delta_{\ell+1} - \delta_\ell)~.
\label{eq:sigmatr}
\end{eqnarray}

In order to calculate phase shift, the Schrodinger equation for the reduced DM two-particles system needs to be solved. In terms of wave-function, this reduced system is expressed as: $\psi(r)=\Sigma_{\ell,m}R_{\ell}(r)Y_{\ell,m}(\theta,\phi)$. The solution for the radial part of the wavefunction, $R_{\ell}(r)$ gives the phase shift parameter, $\delta_{\ell}$. The general form of this equation is given as:
\begin{equation}
\frac{1}{r^2} \frac{d}{dr} \Big( r^2 \frac{d R_{\ell}}{dr} \Big) + \Big( k^2 - \frac{\ell (\ell + 1)}{r^2} - 2\mu V(r) \Big) R_\ell = 0 \label{radial}
\end{equation}
Here, the reduced mass $\mu = m_X/2$ and the momentum $k = \mu v$. This equation~\ref{radial} can be modified by performing certain change of variables. These modifications help to reduce the equation in a compact form without losing the relevant information. Following the conventions of reference~\cite{Tulin:2013teo}, the defined variables are:
\begin{equation}
\chi_\ell \equiv r R_\ell\, , \quad x \equiv \alpha_X m_X r \, , \quad a \equiv \frac{v}{2\alpha_X}\, , \quad b \equiv \frac{\alpha_X m_X}{m_\phi}. \label{vardefs}
\end{equation}
The modified Schrodinger equation~\ref{radial} is given as:
\begin{equation} \label{radial2}
\left( \frac{d^2 }{d x^2} +  a^2 - \frac{\ell(\ell+1)}{x^2} \pm \frac{1}{x} \, e^{-x/b} \right) \chi_\ell(x) = 0 \; .
\end{equation}

\noindent References~\cite{Buckley:2009in} and~\cite{Tulin:2013teo} have followed two different approaches to solve equation~\ref{radial2}. In~\cite{Buckley:2009in}, authors assume the boundary conditions that $\chi_{\ell}$ is regular at origin and $\chi_{\ell}(x)\rightarrow$ $\alpha m_{\chi}(\frac{v}{2\alpha}x-\frac{\pi \ell}{2}+\delta_{\ell})$ at large r values. In~\cite{Tulin:2013teo}, authors follow a more rigorous approach and solve the equation within a fixed domain, $x_i\leq x\leq x_m$. The initial conditions are defined as $\chi_{\ell}(x_i)=1$ and $\chi_{\ell}'(x_i)=(\ell+1)/x_i$. The point $x_i$ is taken to be close to the origin, whereas $x_m$ is chosen in such a way that the potential term in the Schrodinger equation is suppressed compared to the kinetic term i.e. $a^2\gg exp(-x_m/b)/x_m$. For the purpose of this work, we have followed the approach used in work~\cite{Tulin:2013teo}. We first choose the initial values of $x_i$ and $x_m$ based on the following conditions:
\begin{equation}
 a^2\gg exp(-x_m/b)/x_m\text{~~and~~}x_i\ll b
\end{equation}
and then decrease(increase) $x_i$($x_m$) until the percentage change in the obtained value of $\delta_{\ell}$ drops down to less than 1\%. Also, the obtained values of $x_i$ and $x_m$ for s-wave mode is used as initial conditions for $\ell>0$ cases. This deiviates our methodology from the method used in reference~\cite{Tulin:2013teo}. Using this, we show that in order to reach the desired values of transfer cross-section, the needed $\ell$ modes are less than 200 compared with the $\sim1000$ modes in the previous work. This can make the program more effective by around 5 times. The same can be seen for a particular parameter-space point from the figure~\ref{fig:lmode}. For the same values of parameters, the saturation to required transfer cross-section value occurs at around 1200 $\ell$ modes in~\cite{Tulin:2013teo}. 

Equation~\ref{radial2} is solved using partial wave analysis to obtain $\sigma_T/m_{\chi}$ for a range of mediator mass, $m_{Z'}$. The obtained plots with coupling values($\alpha_{\chi}=10^{-2}$) and $10^{-3}$ for an attractive and repulsive-type Yukawa potential are shown in figure~\ref{fig:sigmamphi}. From the plots, it can be seen that numerical results matches well with the results from analytical formualae in their respctive regimes and also predicts the behaviour of $\sigma_T$ in the parameter spcae, where the analytical results are not valid. Within the "resonant-regime", quantum effects are evident. This happens due to the exchange of virtual particles many times between the two DM particles before the actual interaction takes place. In DM literature, this effect is known as the \textbf{"Sommerfeld Effect"} illustrated in the figure~\ref{fig:sommerfeld}. This effect enhances the parameter space for $\sigma_T$ vs. $m_{Z'}$ due to the presence of bound states in the case of an attractive potential. Resonant peaks occur where the phase shift($\delta_{\ell})\rightarrow \frac{\pi}{2}$ and anti-resonant valleys form where the phase shift($\delta_{\ell})\rightarrow 0$. As there is no bound state in a repulsive potential, this resonant behaviour is missing there. But still, the numerical results matches well with the analytical results in their respective regimes for this case too.
\subsection{Velocity dependence of scattering cross-section}
In the previous sub-section, we studied the dependence of scattering cross-section on the mediator mass using the standard Schrodinger equation in the presence of a Yukawa type potential. The motivation for choosing a Yukawa type potential between dark matter and the mediator comes from the fact that resulting scattering cross-section has a strong velocity-dependence, yielding negligible cross-sections at large scales (because of high DM velocities) and enhanced cross-sections at sub-galactic scales(due to low DM velocities). The velocity dependence of scattering cross-section calculated numerically by performing a partial wave analysis can be seen in figure~\ref{fig:velnum}. 

\begin{figure*}[t!]
\begin{center}
\includegraphics[scale=0.9]{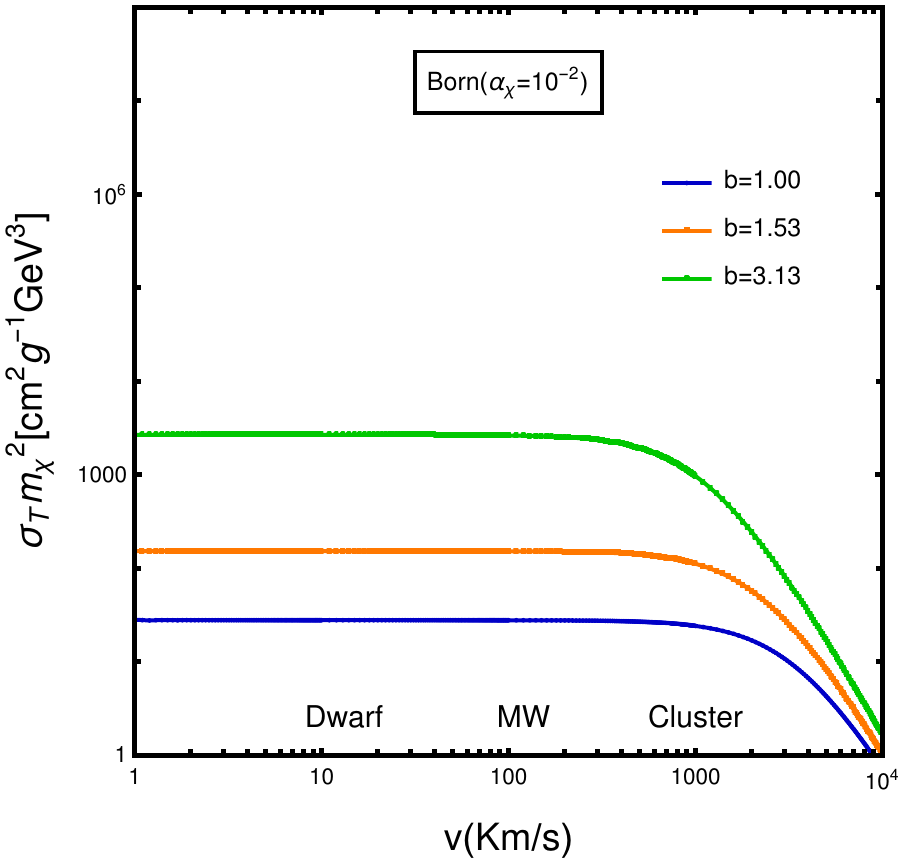}
\includegraphics[scale=0.9]{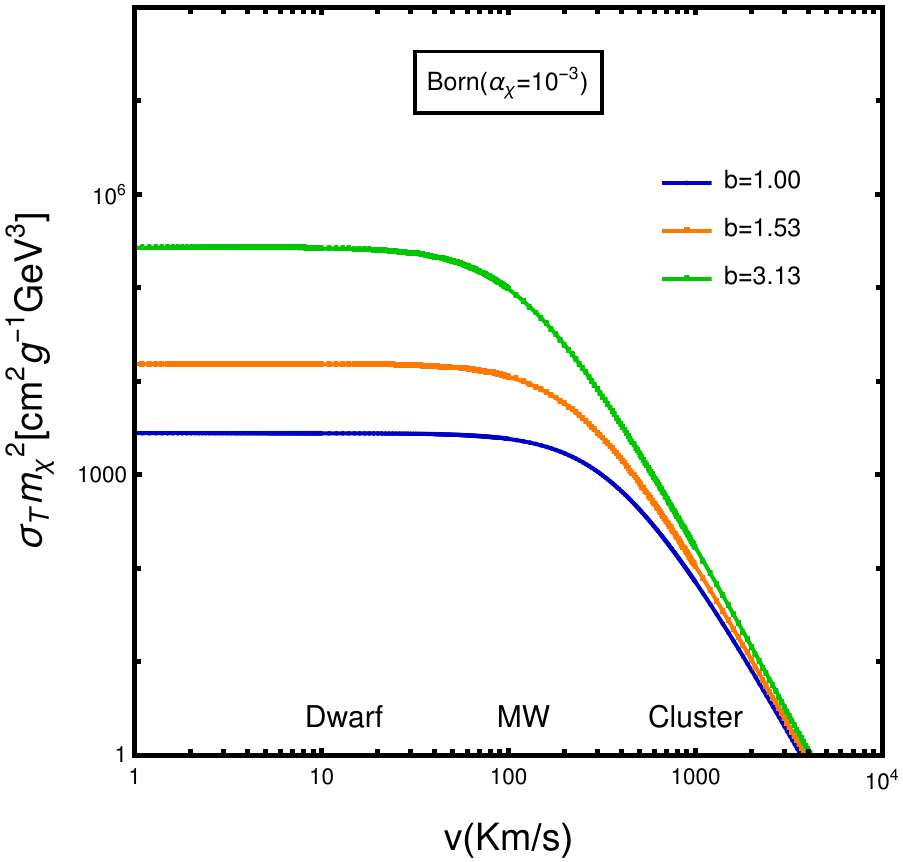}
\includegraphics[scale=0.9]{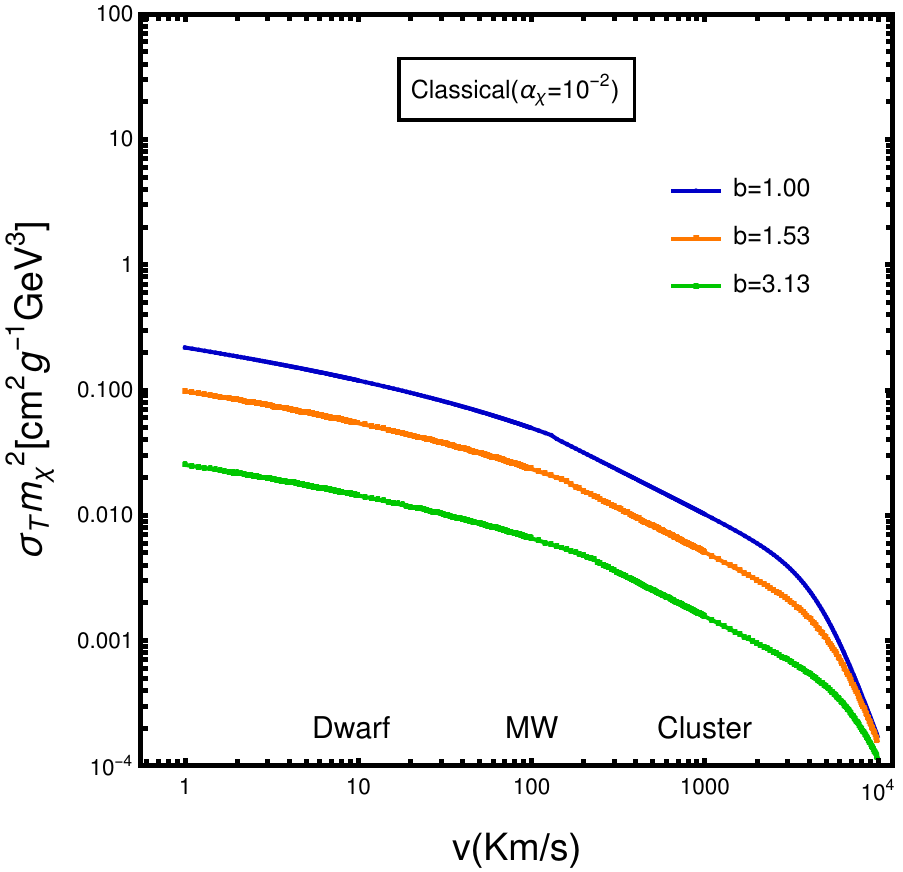}
\includegraphics[scale=0.9]{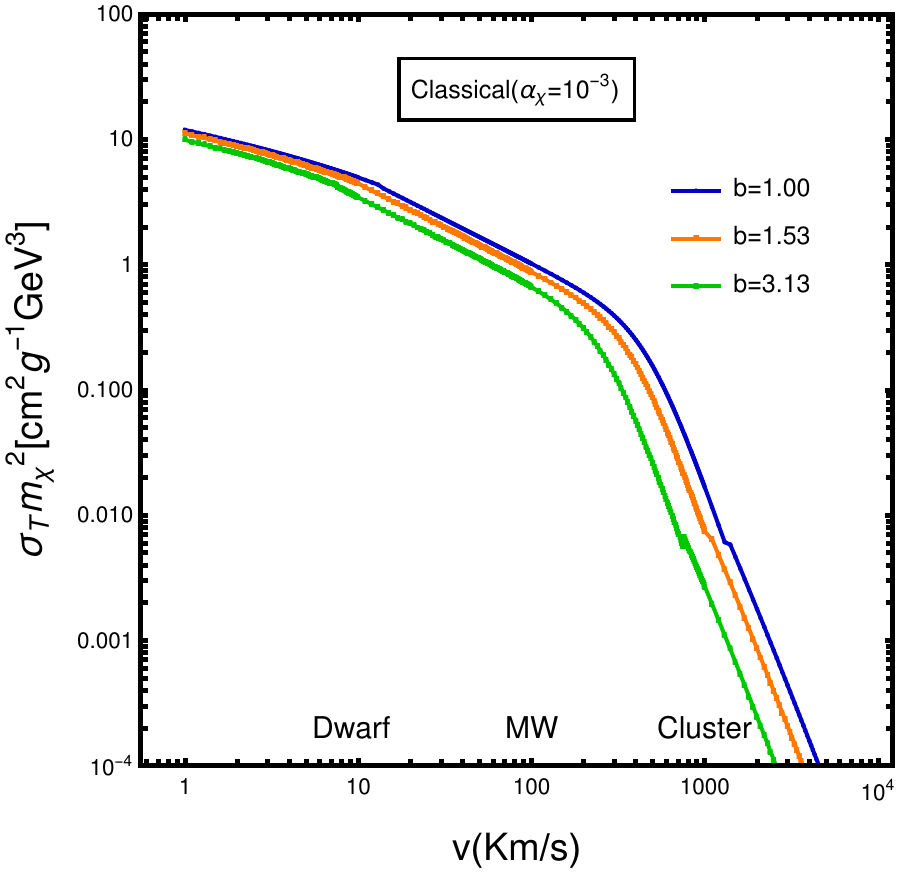}
\end{center}
\caption{Velocity dependence of $\sigma_T$ for different values of coupling constants. Each subimage contains 3 plots each corresponding to 3 different values of parameter "b". We see that both the classical and born formulae predicts a non-dependence of cross-section on velocity of DM at large scales and the cross-section is supressed to negligible values reflecting a non self-interacting nature of DM particles at large scales.}
\label{fig:vel}
\end{figure*}

The dynamics here is quite complicated. There is no clear relation of scattering cross-section on velocity at small scales for different "b" values because of the presence of phase factor in the formula for cross-section. When for a specific value of "b", the phase factor $\delta\rightarrow \frac{\pi}{2}$, then the cross-section is enhanced, as in the case for "b=1.68", similarly when the phase factor $\delta\rightarrow 0$, then the cross-section is strongly suppressed, as can be seen for "b=4.52"  But as the velocity of DM particles increase, we see that dependence of cross-section on the ratio of DM to mediator mass($m_{\chi}/m_{Z'}$) no longer exists, and the value of $\sigma_T$ converge to the Coulomb result, $\sigma_T m^2_{\chi} \propto v^{-4}$. This is a desirable result and is the main motivation for choosing the Yukawa potential for the self-interactions among DM particles.
\par The rich behaviour of numerical results including the resonance zones are not present in the analytical classical and born results. Nevertheless, the overall behaviour of cross-section with the DM velocity can be verified from those results. The same has been provided in the figure~\ref{fig:vel}. We clearly see that with the increase in DM velocity, the scattering cross-section is suppressed for all cases irrespective of the ratio of DM to mediator mass($m_{\chi}/m_{Z'}$) as pointed in the numerical results.   

\begin{figure*}[t!]
\begin{center}
\includegraphics[trim={2.0cm 0 1.5cm 0},clip,scale=0.75]{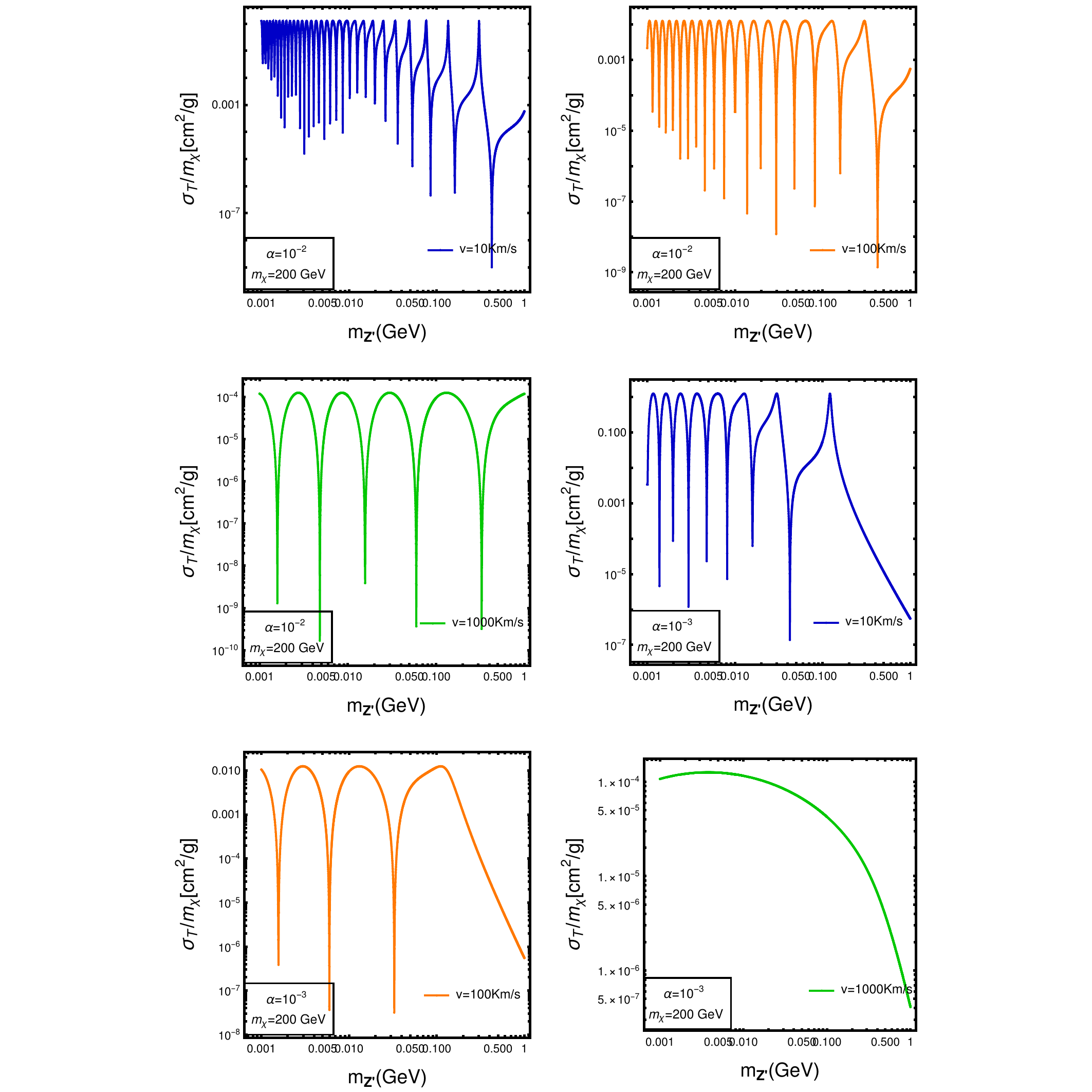}
\end{center}
\caption{Plots for scattering cross-section($\sigma_T/m_{\chi}$) for different DM velocity values with respect to the mediator mass($m_{Z'}$).The value of coupling($\alpha_{\chi}$) is set to $10^{-2}$ for the first 3 plots and $10^{-3}$ for the remaining 3 plots.. The dark matter mass is chosen to be 200GeV and the potential used here is Hulth\'{e}n Potential.}
\label{fig:halthensigma1}
\end{figure*}

\section{A proxy potential with analytical solution}
The resonant region can be explored analytically using a potential similar in it's behaviour and form with the yukawa potential. This potential is called as the Hulth\'{e}n Potential.  Although, it is to be noted that Hulth\'{e}n Potential is analytically solvable for $\ell=0$ i.e. s-wave scattering only. But, nevertheless it can be used as a useful subtitute in the resonant, non-perturbative regime. The form of Halth\'{e}n potential is given as:
\begin{equation} \label{hulthpot}
V(r) = \pm  \frac{\alpha_{\chi} \delta \, e^{- \delta r}}{1 - e^{-\delta r}} \; ,
\end{equation}

The analytical expression for self-interaction transfer cross-section in the presence of an hulth\'{e}n potential in the s-wave limit is given as:
\begin{equation} \label{eq:hulthcross}
 \sigma_T^{\text{Hulth}}=\frac{16\pi\sin^2\delta_0}{m_{\chi}^2v^2}
\end{equation}
Here, the value of $\delta_0$ is given by:
\begin{multline}
\delta_0 ={\rm arg} \Bigg(\frac{i\Gamma \Big(\frac{i m_\chi v}{k M_{Z'}}\Big)}{\Gamma (\lambda_+)\Gamma (\lambda_-)}\Bigg)\\
\lambda_{\pm} = \left\{
			\begin{array}{l}
				1+ \frac{i m_\chi v}{2 k M_{Z'}}  \pm \sqrt{\frac{\alpha_D m_\chi}{k M_{Z'}} - \frac{ m^2_\chi v^2}{4 k^2 M^2_{Z'}}} ~~~~ {\rm Attractive}\\
				1+ \frac{i m_\chi v}{2 k M_{Z'}}  \pm i\sqrt{\frac{\alpha_D m_\chi}{k M_{Z'}} + \frac{ m^2_\chi v^2}{4 k^2 M^2_{Z'}}} ~~~~ {\rm Repulsive}\\
			\end{array}
			\right.
\end{multline}  
\begin{figure*}[t!]
\begin{center}
\includegraphics[trim={3.7cm 0 3.45cm 0},clip,scale=0.75]{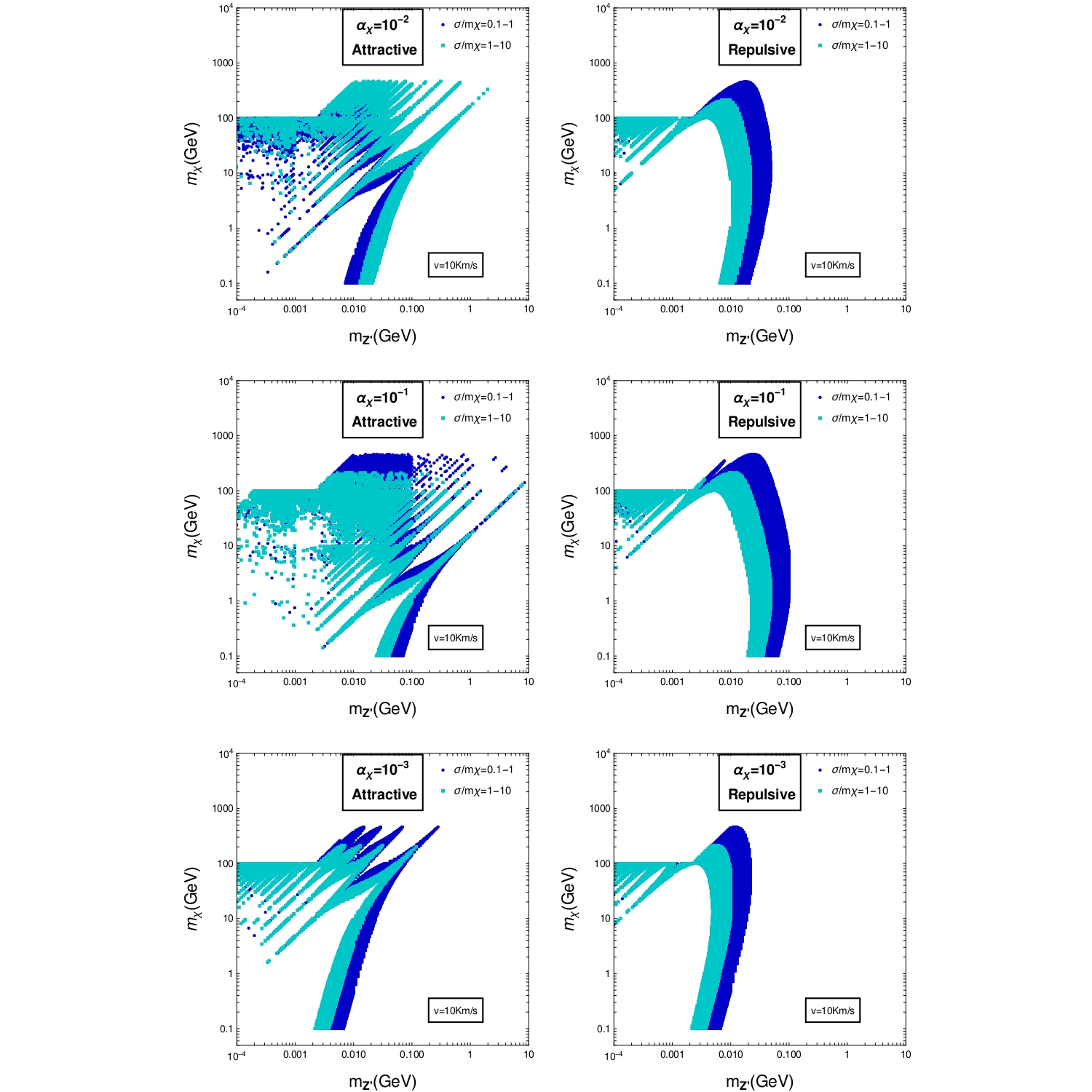}
\end{center}
\caption{Parameter Space for SIDM. The plots has been made using the analytical formula available for Hulth\'{e}n potential. The chosen cross-section($\sigma_T/m_{\chi}$) values are between $0.1~10$, which have been found to solve small scale crisis.}
\label{fig:parameter}
\end{figure*}
\begin{figure*}[t!]
\begin{center}
\includegraphics[scale=0.8]{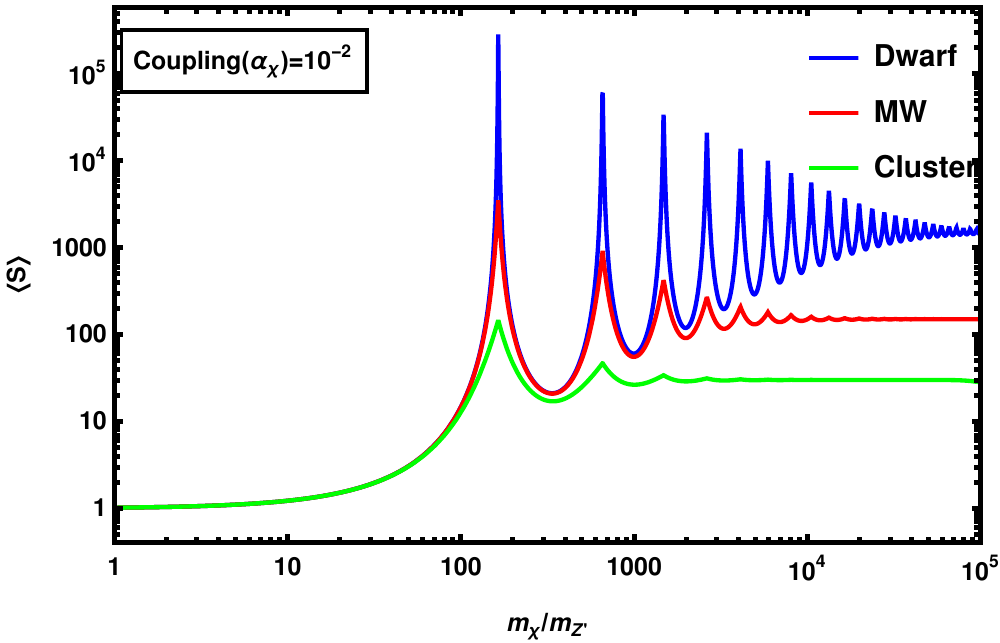}
\includegraphics[scale=0.8]{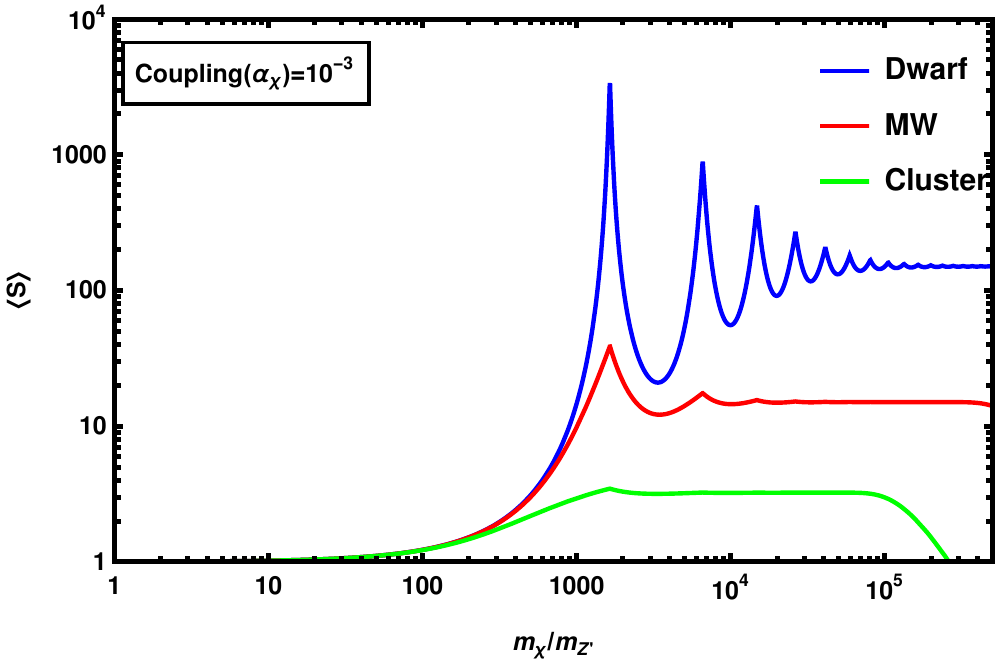}\\
\hspace{-0.15in}\includegraphics[scale=0.8]{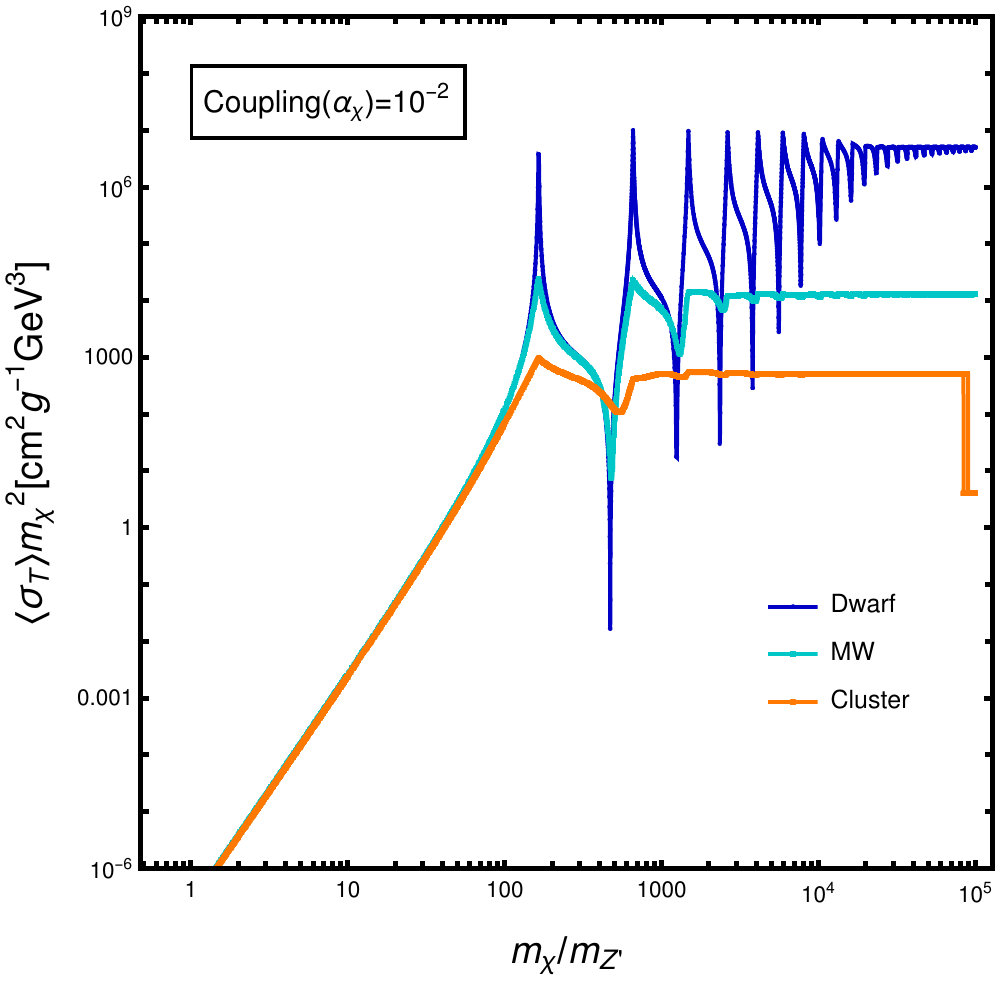}
\includegraphics[scale=0.8]{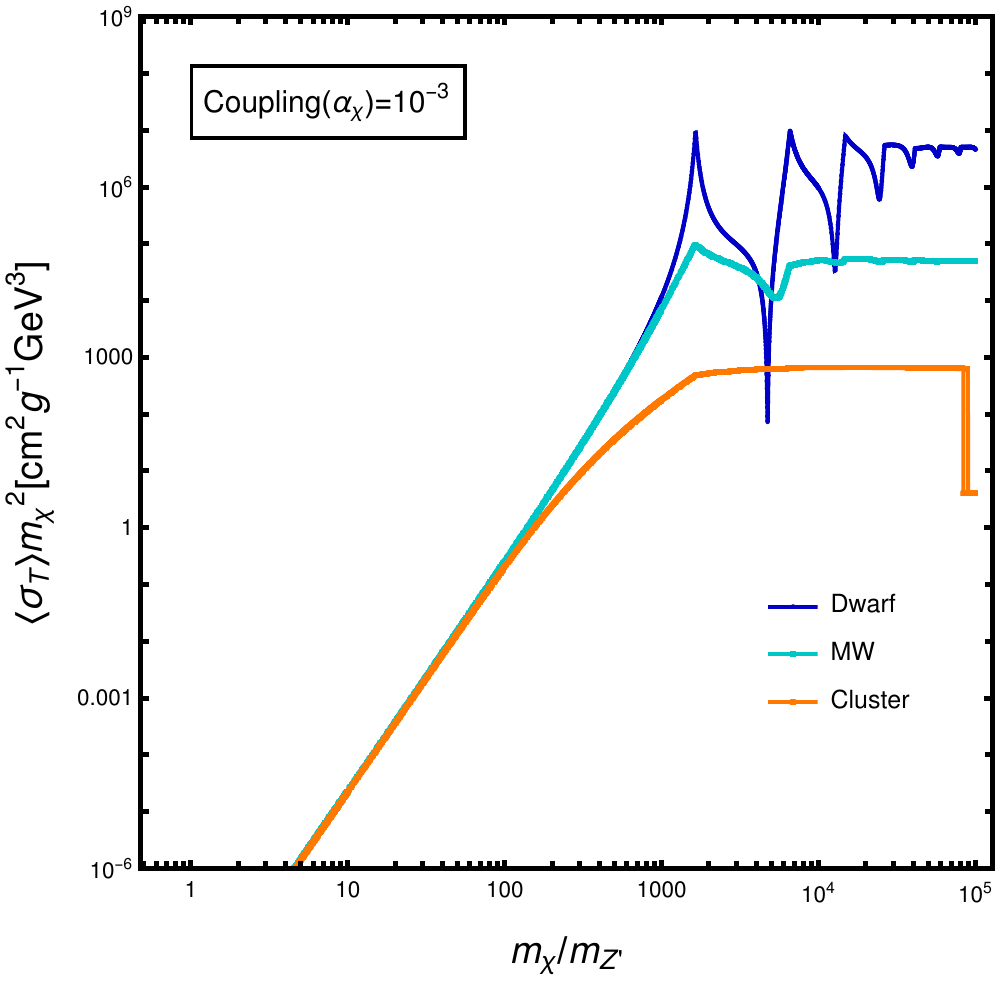}
\end{center}
\caption{This figure shows the dependence of thermally-averaged sommerfeld factor on the ratio of DM to mediator mass in the upper half. We see a strong resonance features for a ratio of around 100-10000 in both the $\alpha =10^{-2} \text{ and } 10^{-3}$ cases. For a comparison, the dependence of thermally-averaged scattering cross-section on the same ratio has been plotted in the lower half of the image. Thus, we can see a clear co-relation between the enhancements inthe thermally averaged DM scattering cross-section and DM annihilation cross-section.}
\label{fig:sommerfeld2}
\end{figure*}
The detailed calculation to derive this result can be found in the reference~\cite{Cassel:2009wt}. Now, using this analytical expression~\ref{eq:hulthcross}, we plot the dependence of cross-section on mediator mass and also on DM particle velocity. Plots in figure~\ref{fig:halthensigma1} show the dependence of scattering cross-section on mediator mass at different scales(DM velocities).  The DM mass has been taken as 200 GeV and the coupling($\alpha_{\chi}=10^{-2}$). The results have been obtained for an attractive potential to emphasize on the resonant nature of plots. Although, as this is an approximation to the original yukawa potential, the point to point matching of the results would give varying results but for understanding the overall structure and dependence of $\sigma_T$, it is a useful result. The arguments presented in the numerical calculations regarding the appearance of resonant peaks and anti-resonant valleys can be applied here too. So, wherever the value of $\delta_0 \rightarrow 1$, the appearance of peaks is seen and similarly $\delta_0 \rightarrow 0$ gives rise to valleys.

\subsection{Parameter Space for SIDM}
As it has been pointed out already that a scattering cross-section($\sigma_T/m_{\chi}$) of around $0.1-10 cm^2/g$ is required at the dwarf scale to resolve the small scale structure anomalies. Using this result, the plot~\ref{fig:parameter} has been made for the allowed ranges of dark matter particle mass($m_{\chi}$) and for the mediator particle mass($m_{Z'}$) within these observational constraints for an attractive as well as repulsive potential. For a given parameter choice, we calculate the thermally averaged scattering cross-section($\sigma_T$) using the approximated Hulth\'{e}n potential. As the Hulth\'{e}n potential is only solvable exactly for $\ell$=0 i.e. s-wave mode, so the results are not valid where the higher $\ell$ modes are required to get correct cross-section values. This is marked by the condition $m_{\chi}v/m_{Z'}\geq1$. The results presented in ~\ref{fig:parameter} thus are plotted only for the case where $m_{\chi}v/m_{Z'}\leq1$. The cross-section of around $0.1 cm^2/g$ has been found to be not only solve small-scale crisis but also be consistent with structure formation at large scales~\cite{Rocha:2012jg}.
\par The expression for thermally averaged transfer cross-section is given by~\ref{eq:velavrsigma}.
\begin{multline}
\label{eq:velavrsigma}
 \langle \sigma_T \rangle = \int \frac{ d^3 v_1 d^3 v_2}{(\pi v_0^2)^{3}} \, e^{-v_1^2/v_0^2}  \, e^{-v_2^2/v_0^2} \, \sigma_T(|\vec{v}_1 -\vec{v}_2|)\\ = \int \frac{d^3 v}{(2\pi v_0^2)^{3/2}} \, e^{-\frac{1}{2} v^2/v_0^2} \, \sigma_T(v) \; ,
\end{multline}
Here, $v_0$ is the most probable velocity and $v = |\vec{v}_1 -\vec{v}_2|$ is the relative velocity.  We choose $v_0$ to be characteristic of dwarf scale halo size. Although velocity-averaging is clearly irrelevant for a constant cross section, it is especially important for strongly velocity-dependent cross sections ({\it e.g.}, resonant features, present in the case of an attractive potential). For fixed $\langle\sigma_T\rangle/m_X$, the (anti)resonances favor larger (smaller) $m_X$, corresponding to peaks pointing to the upper right (lower left) in Fig.~\ref{fig:parameter}.

\section{Relic density constraints on SIDM}
For explanation of small scale issues like cusp-core and TBTF problems, the requirement is that the self scattering of SIDM must give large values of cross-section i.e $\sigma_{\chi \chi\to\chi\chi}/m_{\chi} \approx \mbox{0.1-10\,cm}^2/\mbox{g}$ at dwarf scales~\cite{Dave:2000ar,Vogelsberger:2012ku,Rocha:2012jg,Zavala:2012us,Peter:2012jh}. For comparison, this required SIDM cross section is fourteen orders of magnitude greater than the typical WIMP annihilation cross section $\sigma^{}_{\chi \chi\to \mbox{SM}\, \mbox{SM}}/m_{\chi} \approx 10^{-14} \mbox{cm}^2/\mbox{g}$. At large scales, the relic density calculation is dependent on SIDM annihilation channels as well as production mechanism and thereby, can provide tight constraints on model parameters. 

In order to calculate the amount of SIDM abundance starting from early universe epoch till now, it is useful to study the evolution of SIDM comoving number density using relevant production/depletion mechanisms. This would include the interaction of DM particles with radiation and other SM particles. Such calculations are taken care of by the Boltzmann's equation, which for our case would be given as:
\begin{equation}
\label{eqn:relic-dydx}
\frac{d Y_{\chi}}{dx}= - \frac{\lambda(x)}{x^2} \left[ Y^2_{\chi}-Y^2_{\rm eq} \right]
\end{equation}
with the factor $\lambda(x)$ and variable $Y_{eq}$ being expressed as,
\begin{eqnarray}
& & \lambda(x) \equiv \left(\frac{\pi}{45}\right)^{1/2}\,M_{\rm pl} M_{\rm DM}
\left(\frac{g_{*s}}{\sqrt{g_{*}}} \right) \, \langle \sigma v \rangle(x) \\
& &Y_{\rm eq}=\frac{45}{2 \pi^4} \left(\frac{\pi}{8}\right)^{1/2} \frac{g_{\rm DM}}{g_{*s}} x^{3/2} e^{-x}
\end{eqnarray} 
Here $x=m_{\chi}/T$ and $Y_{\chi}=n_{\chi}/s$, with $n_{\chi}$ being the SIDM number density and $s$ as the entropy density. The other factors are $Y_{\rm eq}$ as the equilibrium value of $Y_{\chi}$, $M_{\rm pl} \simeq 1.2 \times 10^{19}~\mbox{GeV}$ is the Planck mass, $\big<\sigma v\big>$ the thermally-averaged SIDM annihilation cross section, and $g_{*s}$ and $g_*$ are the relativistic degrees of freedom.

In the nonrelativistic limit, the thermally-averaged annihilation cross section $\langle \sigma\, v \rangle$ reduces to an average over a Maxwell-Boltzmann distribution function \cite{Kolb:1990vq}: 
\begin{equation}
\label{eqn:sigma}
\langle \sigma\, v \rangle = \frac{x^{3/2}}{2 \pi^{1/2}} \int^1_0  
(\sigma \, v)\, v^2 \, e^{-\frac{x v^2}{4}} dv 
\end{equation}
Here, $v$ is the relative velocity of the annihilating particles.

It is to be pointed that the actual annihilation cross section times velocity is given as:
$(\sigma \,v)= {\rm S} (\sigma \,v)_0$.
Here, S is the Sommerfeld-enhancement factor, which has been previously discussed~\ref{fig:sommerfeld} in the context of self-interactions of DM in a yukawa potential. The plot for thermally averaged S-factor, relevant for DM annihilation calculations is shown in figure~\ref{fig:sommerfeld2}. For the case when the cross section is enhanced by the Sommerfeld factor, the s-wave thermally-averaged annihilation cross section is given by
\begin{equation}
\label{eqn:sigmaSomm}
\langle \sigma\, v \rangle_{\rm Somm} = \frac{x^{3/2}}{2 \pi^{1/2}} \int^1_0  
(\sigma \,v)_0\,{\rm S}\,  v^2 \, e^{-\frac{x v^2}{4}} dv 
\end{equation}

\begin{figure*}[t!]
\includegraphics{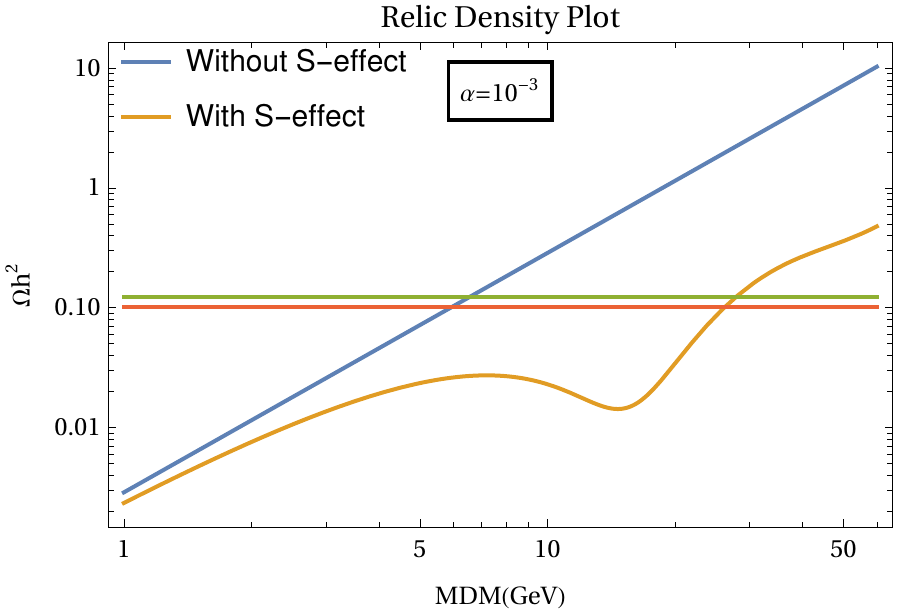}
\caption{The plot shows the obtained relic density as a function of dark matter particle mass in GeV. The bounds obtained on DM relic density from the Planck satellite data~\cite{Planck:2013pxb} has also been plotted(green and red line). The blue line plots the $\Omega h^{2}$ incorporating the Sommerfeld factor in the velocity avearged calculations. We also plot the required $\Omega h^{2}$(orange line) if the Sommerfeld effect is neglected in the early Universe.}
\label{fig:relicden}
\end{figure*}

Here, $S(x)$ is the sommerfeld factor given in equation~\ref{eq:se}.\\
The corrected cross section due to the distortion of incoming states from the plane wave can be calculated from the wave function of the two-body system in the attractive potential of the exchanged light particles. We consider a Dark Matter particle of mass $m_{\chi}$ and the exchanged boson mass $Z'_\mu$ and the coupling between them is $\alpha$. In this paper, we use an analytical approximation for the Sommerfeld factor which for the s-mode can be expressed as \cite{Tulin:2013teo}:
\begin{equation}
S_s=\frac{\pi}{a}\frac{\sinh(2\pi a c)}{\cosh(2\pi ac)-\cos(2\pi\sqrt{c-(ac)^2})},
\label{eq:se}
\end{equation}
Here, $a=v/2\alpha_{\chi}$ and $c=6b/\pi^2 =6\frac{\alpha_{\chi}m_{\chi}}{{\pi}^2m_{Z'}} $

From figure~\ref{fig:sommerfeld2}, it is clear that at large velocity ($v >> \alpha$), there is insignificant enhancement i.e, ${\rm S} \sim 1$. In the intermediate range $v^* < v < \alpha$, the enhancement goes like $1/v$. At very small velocities, a series of resonances appear due to presence of bound states. At the epoch of freeze-out, although the DM particles are non-relativistic, the typical velocities are still very large and in this case, $S(x)$ is assumed to be order of 1 i.e, $\langle \sigma \,v \rangle \sim \langle \sigma \,v \rangle_{\rm Somm}$. But the DM relic density need to be calculated by solving the Boltzmann equation in the late regimes ($t>t_{f}$) where the number density at equilibrium $n_{\rm DM, eq}$ is considerably lower than $n_{\rm DM}$ and hence we can drop $Y_{\rm DM, eq}$ from the Boltzmann equation (\ref{eqn:relic-dydx}). So the freeze-out temperature is defined as the solution of
\begin{equation}
Y(x_f)=(1+c) Y_{eq}(x_f)=(1+c) (0.145)\, \frac{g_\chi}{g_{*s}}\, x_f^{3/2} \,e^{-x_f}\quad;
\end{equation}
where the constant  $c\simeq 1$, $g_{\rm DM}=2$ ; and the freeze-out
temperature is equal to $x_f \simeq 20$. Using these results and approximations, the solution of equation~\ref{eqn:relic-dydx} can be written as: 

\begin{equation}
\frac{1}{Y(x_0)}= \frac{1}{Y(x_f)} + \int_{x_f}^{x_0} dx \frac{\lambda(x) }{x^2}
\label{eqn:Yx0}
\end{equation}

where the limit of integration corresponds to freeze-out $x_f=M_{\rm DM}/T_{f}$ and the 
present day value $x_0=M_{\rm DM}/T_{0}$. By definition, $x_0$ corresponds to the temperature 
where the co-moving density of ${\rm DM}$ does not change noticeably and the integration 
can be terminated. The value of $Y(x_0)$ is connected to the value of the ratio of the Dark 
Matter density to the critical density today ($\Omega_{\rm DM,0}=n_{\rm DM,0} \, M_{\rm DM}/\rho_{\rm crit,0}$):
\begin{equation}
\Omega_{\rm DM,0} h^2=\frac{1}{(8.1\times 10^{-47} \text{GeV}^{4})} M_{\rm DM}\, s(x_0)\,Y(x_0)
\end{equation}
where $s_0= 2918{\rm cm^{-3}}$ is the entropy of the present universe and
$\rho_{\rm crit,0}=h^2 8.1 \times 10^{-47} {\rm GeV^4} $ is the critical density. The
observed value of CDM density from the seven year WMAP data  is $\Omega_c
h^2=0.1123 \pm 0.0105$ (3$\sigma$)~\cite{WMAP:2010qai} (where $h$ is the Hubble parameter in units of $100$ km/s/Mpc). We have solved the Boltzmann equation for DM annihilations numerically in MATHEMATICA for a coupling($\alpha_{\chi}=10^{-3}$) with and without S-effect to demonstrate the effect. The same has been plotted in figure~\ref{fig:relicden}.

\section{Accomodating SIDM in a Leptophilic framework}

Many earlier works have incorporated SIDM in an extension of SM. Firstly, models where SM was extended by a gauge singlet with an interaction to the Higgs were introduced~\cite{Bento:2000ah,McDonald:2001vt}. These models predicted a constant cross-section for self-interactions and also the correct relic abundance if coupled to Higgs boson~\cite{Burgess:2000yq}. Light mediator models with extended $U(1)_D$ symmetry have been extensively studied before~\cite{Loeb:2010gj,Schutz:2014nka,Feng:2009mn,Chen:2006ni,Feldman:2006wd,Carone:2010ha}. Some of the other notable work in the recent literature can be found here~\cite{Foot:2014uba,Foot:2016wvj,Tsai:2020vpi,Koren:2019iuv,Elor:2021swj,Chaffey:2021tmj}. These models provide a way for DM to be charged under these spontaneously broken $U(1)$ symmetry and thus, the conservation of charge automatically ensures DM stability. Also, the mediator for these charges induce self-interactions among DM. The motivation for this section comes from the works described above. Based on the similar line of work, we propose a simple $U(1)_{\ell}$ extension of the standard model which can provide an SIDM and its mediator candidate.

In earlier works~\cite{FileviezPerez:2010gw,Chao:2010mp}, three copies of right-handed neutrinos are introduced along with other exotic fermions in order to make the gauge theory of leptons, anomaly-free. Instead we consider here four different types of chiral fermions $\chi_{iL}$ and $\chi_{iR}$ with $i=1,2$ instead of three copies of right-handed neutrinos with lepton charge as $+1$. The main motivation behind this is two-fold: 
\begin{itemize}
 \item The introduced chiral fermions exactly cancel out the gauge anomalies in the model.
 \item Four chiral fermions give two Dirac fermions(after mass matrix diagonalization) in which lightest one is a stable dark matter candidate. Stability of the dark matter is ensured by the lepton charges of these fields forbidding any decay of the lightest component.
\end{itemize}

The SM scalar Higgs $H$ and gauge singlet scalar $S$ are responsible for the spontaneous symmetry breaking in the theory at the two different required energy scales. $S$ is also required for the DM phenomenology and for giving masses to exotic fermions. The $U(1)_{\ell}$ gauge boson $Z'$ acts as a mediator for self interactions of DM in the case of vector-portal coupling whereas self-interactions can also be induced via scalar $S$ in a scalar-portal type coupling scenario. The particle content of this leptophilic model and other relevant details including the scalar potential and the fermionic mass matrix are given in a tabular structure in the appendix as~\ref{table:1}. We have refrained ourselves from diving into the detailed methodology of this model and its application in the SIDM framework because the same task is being taken up in a follow-up work of ours. In the next section, we comment on the possible parameter space allowed for a SIDM candidate in a leptophilic extended setup of SM from the astrophysical, cosmological and collider data.

\section{Constraints on leptophilic DM candidate}
In this section, we comment on the viable parameter space allowed for a Dirac type SIDM candidate where DM self-interactions are mediated via $U(1)_{\ell}$ vector boson($Z'$). As a $U(1)_{L}$ gauge boson couples both to the DM candidate and to the electrons, so the major constraints on such a framework usually come from direct-detection experiments like XENON1T~\cite{XENON:2018voc}, DAMIC-1K~\cite{Battaglieri:2017aum} and others. Keeping the generality intact, we allow for different values for the coupling of $Z'$ to the SM $U(1)_{\ell}$ current ($g_{\ell}$) and to the dark matter($y_\chi$). 

\subsection{Constraints from Astrophysical and Terrestrial data}

The dominating emission from the celestial bodies like sun, HB, and RG stars, is of electrons but the effect of the nucleon coupling is weak, becoming significant only for $m_{Z'}\lesssim 10^{-2}$ eV~\cite{Hardy:2016kme}. In addition to these constraints, DM capture and annihilation leading to detectable neutrino signals in the sun can also be considered~\cite{Kopp:2010su}, as it is quite natural to assume interaction of DM particles with other leptons as well. Even without any assumptions, the DM annihilation into neutrinos can be generated at loop levels too via W/Z mediation~\cite{Bell:2010ei}. The constraints on such a leptophilic DM via neutrino detection can thus be derived from Super-Kamiokande data~\cite{Super-Kamiokande:2004pou}. While SN1987A constraints have not been derived for the $U(1)_{L}$ case, we can obtain approximate constraints by analyzing previous results in the literature. In the weak coupling regime, the limits on dark photons can be derived from Ref.~\cite{Chang:2016ntp}, this approximates the bounds due to the coupling of the $U(1)_{L}$ gauge boson with electrons. Also, the constraints for trapping limit can be derived by combining the trapping due to absorption from   Ref.~\cite{Rrapaj:2015wgs} and the trapping due to decay of $Z' \to e^+ e^-$ from Ref.~\cite{Chang:2016ntp}.

A branching ratio of 2/5 is obtained for the interaction $Z'\rightarrow e^+e^-$, when the mass of Majorana neutrinos and $Z'$ is below the muon threshold. Thus, the invisible decay $Z'\rightarrow \nu\nu$ can be accounted via the constraints obtained from beam dump \cite{Bjorken:2009mm,PhysRevD.86.095019} and BaBar \cite{Lees:2014xha} for a dark photon case. As compared to the scalar portal only leptophilic interaction of DM, the beam dump constraints are significant even below the $2m_e$ due to the presence of the $Z'\rightarrow\gamma\gamma$ mode. The constraint on $(g-2)_e$ in such a framework can be referred from Ref.~\cite{Davoudiasl:2014kua}. Other than these, there may arise certain additional laboratory level constraints due to $\nu-e$ scattering, and other atomic physics probes althought they may be skipped since these do not change the conclusions; for a summary of these limits as well as new bounds derived from isotope shift measurements, see Refs.~\cite{Frugiuele:2016rii,Delaunay:2017dku}. 
\subsection{Other Direct Detection constraints}
The $U(1)_{\ell}$ gauge boson $Z'$, via the axial current can couple to the active neutrinos. Thus, an interaction like $Z'\to \nu\nu$ decay may sustain the equilibrium of $Z'$ with the neutrinos after their decoupling from the electron-photon plasma. These conclusions when fitted with the comparison of the deuterium abundance \cite{Cyburt:2015mya,Cooke:2013cba} with the predictions in \cite{Boehm:2013jpa} excludes the parameter space where $m_{Z'}\lesssim 10$ MeV. Although for $m_{Z'}\gtrsim10$ MeV, the equilibrium of the neutrinos with the electrons is possible through off-shell $Z'$ exchange which also increases $Neff$. The dominant process in this case is $e^-\nu\rightarrow e^-\nu$ scattering with a cross section of 
\begin{equation}
\sigma_{e^-\nu\rightarrow e^-\nu}\approx\frac{g^4_{L}}{6 \pi}\frac{s}{m_{A'}^4 },
\end{equation}
which requires the thermal averaged rate at $T\approx1$ MeV to be smaller than the Hubble expansion(H). For a detailed discussion including the effect of plasma corrections, \cite{Heeck:2014zfa} can be referred in the context of a $U(1)_{\ell}$ boson interaction with Dirac neutrinos. At the detector level, a leptophilic DM candidate may yield $ej+\cancel{E}$ and $\mu j+\cancel{E}$ signals that can be potentially observed through simple  missing-energy cuts that suppress the Standard Model background at experiments like LHcP~\cite{LHeCStudyGroup:2012zhm} or at future colliders like~\cite{ILC:2013jhg}. Also, a leptophilic DM can be detected at DAMA~\cite{DAMA:2008bis}, since it accepts all types of recoils and extracts the DM signals through its characteristic annual modulation~\cite{Fox:2008kb}. For a Sommmerfeld enhanced DM interaction, due to the small velocity dispersion, a relatively larger annihilation signal in $\omega-$Cen cluster is expected~\cite{Chan:2022amt}. Also, the constraints on a decaying leptophilic DM candidate having mass in sub-GeV to TeV range from the cosmological data have been discussed in ~\cite{Bulbul:2014sua,Boyarsky:2014ska,Essig:2013goa,Cirelli:2020bpc}.      

In the heavy mediator regime, where both the dark matter and the mediator thermalize with the SM, $m_{\chi,Z'}\gtrsim 10$ MeV is allowed by BBN for reasonably large coupling.
For the light mediator regime, it is shown in reference~\cite{Knapen:2017xzo}, that if the entire Dark Matter is self-interacting, none of the proposed target experiments may have the sensitivity once the SIDM constraints are accounted for. If $\chi$ is a subcomponent of the dark matter, such that the SIDM constraints can be relaxed, then there is some accessible parameter space with superfluid helium or Dirac materials.

\section{Conclusion}
In this work, we performed a through numerical analysis of a vector portal SIDM with a light mediator in a model independent approach. We found a rich dynamics of transfer cross-section due to the presence of resonance effect in a attractive yukawa potential. These results were also verified with analytical results obtained using Born and Classical expressions. Our numerical approach was shown to achieve the desired results using lesser angular($\ell$)-modes compared to the previous studies, which is the distinct result for our analysis. Also, our analysis took into account a wider range of dark matter-mediator coupling$\alpha_{\chi}$ by incorporating $\alpha_{\chi}=0.001$ in our calculations, apart from the previously studied $10{-2}$ case.

\par A velocity dependent cross-section, as required for SIDM was also studied in a broad range of parameter space using the numerical analysis. The important parameter is the factor accounting the ratio between dark matter and mediator mass multiplied by the coupling, denoted by "b". It was shown that the Sommerfeld effects on SIDM scattering enhances the transfer cross-section due to the presence of external potential arising from the dark matter-mediator interaction. A parameter space was sketched for the dark matter and the light mediator masses in accordance with the results from various observational results(which require the $\sigma_T/m_{\chi}$ values between $0.1-10 cm^2/g$ for solving the issues at small scales). We found the bounds on $m_{\chi}-m_{Z'}$ plane as $0.1\mbox{GeV}<m_{\chi}<500\mbox{GeV}$ and $0.1\mbox{MeV}<m_{Z'}<0.5\mbox{GeV}$, which have been derived analytically using Hulth\'{e}n potential. The effects of Sommerfeld factor on relic density values were also studied. 

\par Finally, we presented a gauge theory of leptophilic force to accomodate SIDM and its stability alongwith the constrains from various fields (astrophysics, cosmology \& colliders) that decides the allowed parameter space for a SIDM candidate in the leptophilic framework. A detailed study on leptophilic model to incorporate SIDM is left for a future work of ours.

\section{Acknowledgement}
One of the author, Utkarsh Patel would like to acknowledge the financial support obtained from Ministry of Education, Government of India. We would also like to thank Dr. S. Singirala for his useful discussions at the initial stage of the work.

\newpage
\appendix
\section{Gauge theory of leptophilic forces}
The gauge theory of leptophilic forces is based on the following gauge groups
\begin{equation}
 G_{LF} =SU(3)_C \times SU(2)_L \times U(1)_Y \times U(1)_{\ell}\,. 
\end{equation} 
The SM fields and their transformations under this gauge group are presented in table~\ref{table:1}.
\begin{table}[htb!]
\begin{center}
\begin{tabular}{|c|c|}
	\hline
Field	& $ SU(3)_C \times SU(2)_L\times U(1)_Y \times U(1)_{\ell}$	\\
	\hline
	\hline
$Q_{L} =  \begin{pmatrix}
             u_L \\ d_L 
            \end{pmatrix}$	& $\textbf{3} \hspace*{1cm} \textbf{2} \hspace*{1cm}1/6 \hspace*{1cm}0$	\\
 $u_R$	& $\textbf{3} \hspace*{1cm} \textbf{1} \hspace*{1cm}2/3 \hspace*{1cm}0$	\\
 $d_R$	& $\textbf{3} \hspace*{1cm} \textbf{1} \hspace*{1cm}-1/3 \hspace*{1cm}0$	\\
$\ell_L =  \begin{pmatrix}
             \nu_L \\ e_L 
            \end{pmatrix}$	& $\textbf{1} \hspace*{1cm} \textbf{1} \hspace*{1cm}-1/2 \hspace*{1cm}1$	\\
$e_R$	& $\textbf{1} \hspace*{1cm} \textbf{1} \hspace*{1cm}-1\hspace*{1cm}1$	\\
			\hline \hline
\end{tabular}
\end{center}
\vspace{-0.17in}
\caption{SM quarks and leptons with their transformations under the gauge theory of leptons with gauge symmetry $SU(3)_C \times SU(2)_L\times U(1)_Y \times U(1)_{\ell}$. Here, the generation index for all fermions has been omitted.} 
\label{table:1}
\end{table}

A consistent quantum theory must be free from gauge anomalies. So, while introducing a new framework, cancellation of these anomalies needs to be ensured. 
For our $U(1)_{\ell}$ model, the newly introduced gauge anomalies are listed as:
\begin{gather*}
\mathcal{A}[SU(3)^2_C \times U(1)_\ell]  \hspace{.2in} \mathcal{A}[SU(2)^2_L \times U(1)_\ell] \\
\mathcal{A}[U(1)^2_Y \times U(1)_\ell]  \hspace{.2in} \mathcal{A}[U(1)_Y \times U(1)^2_\ell]\\
\mathcal{A}[U(1)^2_Y \times U(1)_\ell]  \hspace{.2in} \mathcal{A}[U(1)_Y \times U(1)^2_\ell]
\end{gather*}

Out of these, the vanishing anomalies are:
\begin{equation*}
 \mathcal{A}[SU(3)^2_C \times U(1)_{\ell}],\hspace{0.3in}\mathcal{A}[U(1)_Y \times U(1)^2_{\ell}],
\end{equation*} 
and the non-vanishing gauge anomalies are given as:
\begin{align*}
& \mathcal{A}[SU(2)^2_L \otimes U(1)_\ell] =\frac{3}{2}, & \mathcal{A}[U(1)^3_\ell] =3, \\
& \mathcal{A}[U(1)^2_Y \otimes U(1)_\ell] =-\frac{3}{2}, & \mathcal{A}[\text{gravity}^2 \otimes U(1)_\ell] =3.
\end{align*}

The generated anomalies can be cancelled by adding a pair of $SU(2)_L$ doublet fermions $(\Psi_{L} \text{ and } \Psi_{R})$, a left-handed isospin triplet $\Sigma_{L}$, a left-handed particles singlet under SM $\rho_{L}$ as presented in the table~\ref{table:2}.
\begin{table}
\begin{tabular}{cccc}
\hline\hline
 & $\text{SU(2)}_L$ & $\text{U(1)}_Y$  & $U(1)_{\ell}$\\
 Gauge fields & $\vec{W}_\mu$ & $B_\mu$ & $Z'_{\mu}$\\[1mm]
\hline Fermions \\ 
(Other than SM) \\
\hline
$\nu_{Ri}(i=1,2,3) $ & $\mathbf{1}$ & $0$  & $1$ \\
$\Psi_L = \begin{pmatrix}
           \Psi^+_L \\ \Psi^0_L
          \end{pmatrix}
$ & $\mathbf{2}$ & $\phantom{+}1/2$  & $\phantom{+}3/2$\\
$\Psi_R = \begin{pmatrix}
           \Psi^+_R \\ \Psi^0_R
          \end{pmatrix}$ & $\mathbf{2}$ & $1/2$  & $-3/2$\\
$\Sigma_L =\frac{1}{\sqrt{2}} \begin{pmatrix}
           \Sigma^0_L & \sqrt{2} \Sigma^+_L \\
           \sqrt{2} \Sigma^-_L & - \Sigma^0_L
          \end{pmatrix}$ & $\mathbf{3}$ & $0$ & $-3/2$\\
$\rho_{L}$ & $\mathbf{1}$ & $0$ & $-3/2$ 
\\[1mm] \hline
Scalar fields\\
$H$ & $\mathbf{2}$ & $\phantom{+}1/2$  & $0$ \\
$S$ & $\mathbf{1}$ & $0$ & $3$ \\
\hline\hline
\end{tabular}
\caption{Gauge theory of leptons with field content and their transformations under the gauge group $SU(2)_L \times U(1)_Y \times U(1)_{\ell}$. These fields are chiral in nature and are color singlets while the hypercharge is given by the relation, $Q = T^3 + Y$.}
\label{table:2}
\end{table}

The relevant kinetic Lagrangian for the present framework is given as follows
\begin{align}
\label{eq:TheModel}
\mathcal{L}_{\ell}
&=     i \, \overline{\Psi_{L}} \left( \slashed{\partial} + ig\frac{\sigma^i}{2}W^i_{\mu}+ig'\frac{1}{2}B_{\mu}+\frac{3}{2} i\,g_\text{L}  \,Z_\mu^\prime \gamma^\mu \right)\,\Psi_{L} \nonumber\\
&~~~+ i \, \overline{\Psi_{R}} \left( \slashed{\partial} + ig\frac{\sigma^i}{2}W^i_{\mu}+ig'\frac{1}{2}B_{\mu}-\frac{3}{2} i\,g_\text{L}  \,Z_\mu^\prime \gamma^\mu \right)\,\Psi_{R}
    \nonumber\\
&~~~= i \, \overline{\Psi} \left( \slashed{\partial} + ig\frac{\sigma^i}{2}W^i_{\mu}+ig'\frac{1}{2}B_{\mu}-\frac{3}{2} i\,g_\text{L}  \,Z_\mu^\prime \gamma^\mu \right)\,\Psi
    \nonumber\\
\end{align}

In $U(1)_{\ell}$, the lepton charges for quarks are zero, while for known SM leptons is +1. 
\begin{equation}
 g_{\ell} Z'_{\mu}\sum_l \Bigg(\overline{l_L}\gamma^{\mu}l_L+\overline{l_R}\gamma^{\mu}l_R\Bigg)
\end{equation}
The DM interactions with this $Z'$ is given as follows:
\begin{multline}
\overline{\Psi}i\gamma^{\mu}\Bigg(\partial_{\mu}+ig\frac{\sigma^i}{2}W^i_{\mu}+ig'\frac{1}{2}B_{\mu}  \Bigg)\Psi-m_{\Psi}\overline{\Psi}\Psi\\
+g_{\ell}\overline{\Psi}\gamma^{\mu}(\frac{3}{2}P_{L}-\frac{3}{2}P_{R})\Psi Z'_{\mu}
\end{multline}
The interaction of DM with the Standard Model gauge bosons is given by:
\begin{multline}
 \mathcal{L}_{\Psi}=\Bigg(\frac{e}{2\sin\theta_W\cos\theta_W} \Bigg)\overline{\Psi^0}\gamma^{\mu}Z_{\mu}\Psi^0\\+ \frac{e}{\sqrt{2}\sin\theta_W}\overline{\Psi^0}\gamma^{\mu}W_{\mu}^+\Psi^-\\ + \frac{e}{\sqrt{2}\sin\theta_W}\overline{\Psi^+}\gamma^{\mu}W_{\mu}^-\Psi^0-eN^+\gamma^{\mu}A_{\mu}N^- \\- \Bigg(\frac{e}{2\sin\theta_W\cos\theta_W} \Bigg)\cos 2\theta_W\Psi^+\gamma^{\mu}Z_{\mu}\Psi^{-}
\end{multline}

\section{Constructing the fermion sector}
Inclusion of extra fermions $\Psi_{L}, \Psi_{R}, \Sigma_{L} \text{ and }\rho_{L}$ to Standard Model prvoides a novel possibility of stable DM candidates in the context of SIDM. The lightest neutral fermion will be the DM candidate. There is a possibility of Majorana($\Sigma^0, \rho^0$) as well as a Dirac($\Psi=\Psi_L+\Psi_R$) component of DM. The viability of Majorana Dark Matter $\Sigma^0, \rho^0$ can be studied at a later time, while our main focus of this work is to consider resulting Dirac fermion $\Psi$ as SIDM candidate. In order to establish a Dirac fermion $\Psi$ as the SIDM candidate, we write down the exotic fermions mass matrix as:

\begin{eqnarray}
M^2_{NF}= \left(\begin{array}{cccc|c} \overline{\Psi_1^0}   &   \overline{\Psi_2^0} & \overline{\Sigma^0} & \overline{\rho^0}  \\  \hline
0 & \frac{\text{vB} \lambda \psi }{\sqrt{2}} & -\frac{\text{h3} v}{2 \sqrt{2}} & \frac{\text{h2} v}{\sqrt{2}} & (\Psi_1^0)^c \\
 \frac{\text{vB} \lambda \psi }{\sqrt{2}} & 0 & -\frac{\text{h4} v}{2 \sqrt{2}} & \frac{\text{h1} v}{\sqrt{2}} & (\Psi_2^0)^c\\
 -\frac{\text{h3} v}{2 \sqrt{2}} & -\frac{\text{h4} v}{2 \sqrt{2}} & \frac{\text{vB} \lambda \Sigma }{\sqrt{2}} & 0 & (\Sigma^0)^c \\
 \frac{\text{h2} v}{\sqrt{2}} & \frac{\text{h1} v}{\sqrt{2}} & 0 & \frac{\sqrt{2} \text{vB} \lambda \chi}{1} & (\rho^0)^c  \\
\end{array}\right)\,
\end{eqnarray}

This after the complete diagonalization gives the relevant mass matrix for the framework.

\section{Gauge Boson Masses}
The vevs of the scalar singlet $S$ result in mass to the neutral gauge boson $Z^\prime$, while
spontaneous symmetry breaking of $H$ generates mass terms for the other neutral gauge bosons $B$ and $W^{3}$.
Assuming negligible kinetic mixing among the $U(1)$ gauge bosons, the resulting neutral gauge boson mass matrix  
in the basis $(W_{\mu}^{3},~B_{\mu},~Z_\mu^\prime)$ yields
\begin{align}
\mathcal{M}_\text{neutral}^2
=
\left(\begin{array}{ccc}
\frac{1}{4} g^2 \langle H \rangle^2 & -\frac{1}{4} g g^\prime \langle H \rangle^2 & 0 \\
-\frac{1}{4} g g^\prime \langle H \rangle^2 & \frac{1}{4} g^{\prime^2} \langle H \rangle^2 & 0 \\
0 & 0 & 9 g^2_{\ell} \langle S_{\ell} \rangle^2 \\
\end{array} \right)
.
\end{align}
Hence, besides the SM $Z$ boson with $M_Z=\frac{1}{2} \langle H \rangle \sqrt{g^2+g^{\prime^2}}$ and the massless photon, there is an extra massive gauge boson, $Z^\prime$, with mass
\begin{eqnarray}
M^2_{Z^\prime}
=
9 g^2_{\ell} \langle S_{\ell} \rangle^2
\end{eqnarray}

The neutral gauge boson corresponding to $U(1)_{\ell}$ group is completely decoupled from the other neutral gauge bosons of SM. Thus, the weak and mass eigenstates are same for the leptophilic gauge boson and denoted as $Z'_{\mu}$.

\section{New proposed $U(1)_{\ell}$ anomaly-free models}
\begin{table}
\begin{tabular}{cccc}
\hline\hline
 & $\text{SU(2)}_L$ & $\text{U(1)}_Y$  & $U(1)_{\ell}$\\
 Gauge fields & $\vec{W}_\mu$ & $B_\mu$ & $Z'_{\mu}$\\[1mm]
\hline Fermions \\ 
(Other than SM) \\
\hline
$\zeta_{L}$ & $\mathbf{1}$ & $0$ & $-4/3$\\
$\eta_{L}$ & $\mathbf{1}$ & $0$ & $-1/3$\\
$\chi_{1R}$ & $\mathbf{1}$ & $0$ & $+2/3$\\
$\chi_{2R}$ & $\mathbf{1}$ & $0$ & $+2/3$\\
\\[1mm] \hline
Scalar fields\\
$H$ & $\mathbf{2}$ & $\phantom{+}1/2$  & $0$ \\
$\phi_1$ & $\mathbf{1}$ & $0$ & $1$ \\
$\phi_2$ & $\mathbf{1}$ & $0$ & $2$ \\
\hline\hline
\end{tabular}
\caption{Model A: New variant of gauge theory of leptons with field content and their transformations under the gauge group $SU(2)_L \times U(1)_Y \times U(1)_{\ell}$. Here, we have replaced the right-handed neutrinos $\nu_{Ri}(i=1,2,3)$ from the model in table~\ref{table:2} with these proposed fermions. All the other fermions remain same and thus can be referred from there. Also, the scalar sector has been modified.}
\label{table:3}
\end{table}

\begin{table}
\begin{tabular}{cccc}
\hline\hline
 & $\text{SU(2)}_L$ & $\text{U(1)}_Y$  & $U(1)_{\ell}$\\
 Gauge fields & $\vec{W}_\mu$ & $B_\mu$ & $Z'_{\mu}$\\[1mm]
\hline Fermions \\ 
(Other than SM) \\
\hline
$\zeta_{L}$ & $\mathbf{1}$ & $0$ & $-5$\\
$\chi_{1R}$ & $\mathbf{1}$ & $0$ & $+4$\\
$\chi_{2R}$ & $\mathbf{1}$ & $0$ & $+4$\\
 \hline

\end{tabular}
\caption{Model B: New variant of gauge theory of leptons with field content and their transformations under the gauge group $SU(2)_L \times U(1)_Y \times U(1)_{\ell}$. Here, the scalar fields have been omitted as they are same as the ones in Model A~\ref{table:3}.}
\label{table:4}
\end{table}
\newpage
\bibliographystyle{utphys}
\bibliography{SIDM_leptophilic}
\end{document}